\pdfoutput=1

\documentclass[11pt,twoside,a4paper,cmspaper,final,collab]{cms-tdr}
\def\svnVersion{493919P}\def\cmsCernNoTag{CERN-EP-2018-304}\def\cmsCernDate{\today}\def\cmsMessage{Published in Physics Letters B as \href{http://dx.doi.org/10.1016/j.physletb.2019.03.059}{\doi{10.1016/j.physletb.2019.03.059.}}}

\begin{document}\cmsNoteHeader{HIG-17-028}

\hyphenation{had-ron-i-za-tion}
\hyphenation{cal-or-i-me-ter}
\hyphenation{de-vices}
\RCS$Revision: 493066 $
\RCS$HeadURL: svn+ssh://svn.cern.ch/reps/tdr2/papers/HIG-17-028/trunk/HIG-17-028.tex $
\RCS$Id: HIG-17-028.tex 493066 2019-03-29 13:53:07Z tklijnsm $
\newlength\cmsFigWidth
\ifthenelse{\boolean{cms@external}}{\setlength\cmsFigWidth{0.98\columnwidth}}{\setlength\cmsFigWidth{0.49\textwidth}}
\ifthenelse{\boolean{cms@external}}{\providecommand{\cmsLeft}{upper\xspace}}{\providecommand{\cmsLeft}{left\xspace}}
\ifthenelse{\boolean{cms@external}}{\providecommand{\cmsRight}{lower\xspace}}{\providecommand{\cmsRight}{right\xspace}}
\ifthenelse{\boolean{cms@external}}{\providecommand{\cmsAppend}{\relax}}{\providecommand{\cmsAppend}{Appendix\xspace}}
\newlength\cmsTabSkip\setlength\cmsTabSkip{3pt}
\providecommand{\CL}{CL\xspace}

\providecommand{\cmsTable}[1]{\ifthenelse{\boolean{cms@external}}{#1}{\resizebox{\textwidth}{!}{#1}}}
\providecommand{\cmsTableForced}[1]{\resizebox{\textwidth}{!}{#1}}

\newcommand{\hboson}{\ensuremath{\PH}\xspace}
\newcommand{\bquark}{\ensuremath{\cPqb}\xspace}
\newcommand{\cquark}{\ensuremath{\cPqc}\xspace}
\newcommand{\tquark}{\ensuremath{\cPqt}\xspace}
\newcommand{\zboson}{\ensuremath{\cPZ}\xspace}
\newcommand{\photon}{\ensuremath{\cPgg}\xspace}
\newcommand{\jpsi}{\ensuremath{\cPJgy}\xspace}
\newcommand{\proton}{\ensuremath{\Pp}\xspace}
\newcommand{\taulepton}{\ensuremath{\PGt}\xspace}
\newcommand{\electron}{\ensuremath{\Pe}\xspace}
\newcommand{\muon}{\ensuremath{\Pgm}\xspace}
\newcommand{\muF}{\ensuremath{\mu_\text{F}}\xspace}
\newcommand{\muR}{\ensuremath{\mu_\text{R}}\xspace}
\newcommand{\bb}{\bbbar}
\newcommand{\cc}{\ensuremath{\cquark\overline{\cquark}}\xspace}
\newcommand{\hbb}{\ensuremath{\hboson\to\bb}\xspace}
\newcommand{\hcc}{\ensuremath{\hboson\to\cc}\xspace}
\newcommand{\hgg}{\ensuremath{\hboson\to\photon\photon}\xspace}
\newcommand{\hzz}{\ensuremath{\hboson\to\zboson\zboson}\xspace}
\newcommand{\hzztofourl}{\ensuremath{\hzz^{(\ast)}\to 4\ell}\xspace}
\newcommand{\BRgamgam}{\ensuremath{\mathcal{B}_{\photon\photon}}\xspace}
\newcommand{\BRZZ}{\ensuremath{\mathcal{B}_{\zboson\zboson}}\xspace}
\newcommand{\pth}{\ensuremath{\pt^\hboson}\xspace}
\newcommand{\njets}{\ensuremath{N_\text{jets}} \xspace}
\newcommand{\absy}{\ensuremath{\abs{y_\hboson}} \xspace}
\newcommand{\ptjet}{\ensuremath{\pt^\text{jet}} \xspace}
\newcommand{\mH}{\ensuremath{m_{\hboson}}\xspace}
\newcommand{\msd}{\ensuremath{m_\text{SD}}\xspace}
\newcommand{\ggh}{\ensuremath{\Pg\Pg\hboson}\xspace}
\newcommand{\cg}{\ensuremath{c_\Pg}\xspace}
\newcommand{\kappab}{\ensuremath{\kappa_{\bquark}}\xspace}
\newcommand{\kappac}{\ensuremath{\kappa_{\cquark}}\xspace}
\newcommand{\kappat}{\ensuremath{\kappa_{\tquark}}\xspace}
\newcommand{\pb}{\unit{pb}}
\newcommand{\kappabLeftAsimov}{-1.3}
\newcommand{\kappabRightAsimov}{1.3}
\newcommand{\kappabLeftObserved}{-1.1}
\newcommand{\kappabRightObserved}{1.1}
\newcommand{\kappacLeftAsimov}{-6.1}
\newcommand{\kappacRightAsimov}{6.0}
\newcommand{\kappacLeftObserved}{-4.9}
\newcommand{\kappacRightObserved}{4.8}
\newcommand{\kappabLeftAsimovFLOATINGBRS}{-8.8}
\newcommand{\kappabRightAsimovFLOATINGBRS}{15}
\newcommand{\kappabLeftObservedFLOATINGBRS}{-8.5}
\newcommand{\kappabRightObservedFLOATINGBRS}{18}
\newcommand{\kappacLeftAsimovFLOATINGBRS}{-31}
\newcommand{\kappacRightAsimovFLOATINGBRS}{36}
\newcommand{\kappacLeftObservedFLOATINGBRS}{-33}
\newcommand{\kappacRightObservedFLOATINGBRS}{38}
\newcommand{\kappaframework}{$\kappa$-\hspace{0pt}framework}
\hyphenation{para-metrized para-metri-zation para-metri-zations}

\cmsNoteHeader{HIG-17-028} \title{Measurement and interpretation of differential cross sections for Higgs boson production at $\sqrt{s}=13$\TeV}

\date{\today}

\abstract{Differential Higgs boson ($\PH$) production cross sections are sensitive probes for physics beyond the standard model.
New physics may contribute in the gluon-gluon fusion loop, the dominant Higgs boson production mechanism at the LHC, and manifest itself through deviations from the distributions predicted by the standard model.
Combined spectra for the $\PH\to\cPgg\cPgg$, $\PH\to\cPZ\cPZ$, and $\PH\to\bbbar$ decay channels and the inclusive Higgs boson production cross section are presented, based on proton-proton collision data recorded with the CMS detector at $\sqrt{s}=13$\TeV corresponding to an integrated luminosity of 35.9\fbinv.
The transverse momentum spectrum is used to place limits on the Higgs boson couplings to the top, bottom, and charm quarks, as well as its direct coupling to the gluon field.
No significant deviations from the standard model are observed in any differential distribution.
The measured total cross section is $61.1\pm 6.0\stat\pm 3.7\syst\unit{pb}$, and the precision of the measurement of the differential cross section of the Higgs boson transverse momentum is improved by about 15\% with respect to the $\PH\to\cPgg\cPgg$ channel alone.
}

\hypersetup{pdfauthor={CMS Collaboration},pdftitle={Measurement and interpretation of differential cross sections for Higgs boson production at sqrt(s)=13 TeV},pdfsubject={CMS},pdfkeywords={differential cross sections, combination, Higgs boson coupling modifiers}}

\maketitle

\section{Introduction}
\label{sec:introduction}

The Higgs boson (\hboson), whose existence is predicted by the Brout--Englert--Higgs mechanism~\cite{Higgs:1964pj,Englert:1964et,Guralnik:1964eu}, is responsible for electroweak symmetry breaking in the standard model (SM).
Since the discovery~\cite{Aad:2012tfa,Chatrchyan:2012xdj,Chatrchyan:2013lba} of a particle compatible with the SM Higgs boson at the CERN LHC, extensive effort has been dedicated to the measurement of its properties and couplings.

In this analysis we measure the inclusive and differential cross sections for the production of Higgs bosons.
Compared with inclusive measurements~\cite{Khachatryan:2016vau,Aad:2015zhl,Sirunyan:2018koj}, differential distributions provide extended information on the Higgs boson couplings, which can be extracted by fitting parametrized spectra to a combination of differential cross sections.
When the Higgs boson couplings to quarks and to other bosons are varied with respect to their SM values, distortions of the predicted differential cross section spectra appear, which are particularly pronounced in the transverse momentum ($\pt$) distribution.

A precise measurement of the Higgs boson couplings represents an important test of the SM, as the couplings are sensitive to several SM extensions~\cite{Dimopoulos:1981zb,Witten:1981nf}.
While the couplings to the top ($y_\tquark$) and bottom ($y_\bquark$) quarks are known with fair precision, there is still a relatively large uncertainty in the measurement of the couplings to lighter quarks such as the coupling to the charm quark ($y_\cquark$).
A proof-of-concept study determining limits on the modification of the SM Higgs boson coupling ($y_\cquark^\text{SM}$) to the charm quark, $\kappac = y_\cquark / y_\cquark^\text{SM}$, from the Higgs boson transverse momentum ($\pth$) distribution was performed in Ref.~\cite{Bishara:2016jga}.
Reinterpreting the ATLAS Collaboration measurements in Ref.~\cite{Aad:2015lha}, this analysis yields the overall bounds $\kappac \in [ -16, 18 ]$ at 95\% confidence level (\CL).
Using the same data set, a reinterpretation of a search by the ATLAS Collaboration for the $\PH\to\jpsi\photon$ channel~\cite{Aad:2015sda} yields $\abs{\kappac}<429$ at 95\% \CL~\cite{Koenig:2015pha}.
More recently, studies from the ATLAS Collaboration~\cite{Aaboud:2018txb,Aaboud:2018fhh}, using data collected at $\sqrt{s}=13$\TeV corresponding to an integrated luminosity of $36.1$\fbinv, yield an observed upper limit on the $\PH\to\jpsi$ branching fraction of $3.5 \times 10^{-4}$ at 95\% \CL that is an improvement of about a factor two with respect to the result obtained in Ref.~\cite{Aad:2015sda}, and an observed upper limit on the product of the production cross section and branching fraction $\sigma(\proton\proton\to\zboson\hboson) \mathcal{B}(\hcc)$ of $110$ times the SM value at 95\% \CL.

Both the ATLAS and CMS Collaborations have reported measurements of differential Higgs boson production cross sections at $\sqrt{s}=8$ and $13$\TeV~\cite{Aad:2014lwa,Khachatryan:2015rxa,Aad:2014tca,Khachatryan:2015yvw,Aad:2016lvc,Khachatryan:2016vnn,Aaboud:2018xdt,Sirunyan:2018kta,Aaboud:2017oem,CMS_AN_2016-442,Aaboud:2018ezd}.
The CMS Collaboration has measured differential Higgs boson production cross sections in the $\hgg$~\cite{Sirunyan:2018kta} and $\hzztofourl$ ($\ell = \electron$ or $\muon$)~\cite{CMS_AN_2016-442} decay channels using data recorded by the CMS experiment in 2016 at $\sqrt{s}=13\,$TeV, corresponding to an integrated luminosity of $35.9\fbinv$.
We report measurements of differential cross sections obtained by combining these results.
Additionally, we include a search for the Higgs boson produced with large \pt and decaying to a bottom quark-antiquark ($\bb$) pair~\cite{CMS_AN_2016-366} in the combination of the $\pth$ spectra.
The differential cross sections for the following observables are combined: $\pth$, the Higgs boson rapidity $\absy$, the number of hadronic jets $\njets$, and the transverse momentum of the leading hadronic jet $\ptjet$.

We interpret the $\pth$ spectrum in terms of Higgs boson couplings.
In order to take into account as many degrees of freedom as possible, multiple couplings are varied simultaneously.
We present results obtained by varying simultaneously
(i) the modifier of the Higgs boson coupling to the charm quark $\kappac$ and the bottom quark $\kappab$,
(ii) the modifier of the Higgs boson coupling to the top quark $\kappat$ and the coefficient $\cg$ of the anomalous direct coupling to the gluon field in the heavy top quark mass limit,
and (iii) $\kappat$ and $\kappab$.

The SM production cross sections and decay rates depend on the Higgs boson mass $\mH$.
We assume a Higgs boson mass of $125.09$\GeV for all measurements in this paper, based on the combined ATLAS and CMS measurement using proton-proton collision data collected in 2011 and 2012~\cite{Aad:2015zhl}.

\section{Theoretical predictions}
\label{sec:theory}

Differential cross sections may be used to constrain model parameters.
In the case of Higgs boson production via gluon fusion, the dominant production mode at the LHC, finite quark mass effects and moderate variations to Higgs boson couplings may manifest themselves through distortions of the $\pth$ spectrum.
We interpret the $\pth$ spectrum for gluon fusion in terms of modifications of the couplings of the Higgs boson using two models: one tailored to heavy quarks and thus sensitive to effects at high $\pt$~\cite{Grazzini:2017szg,Grazzini:2016paz}, and the other considering the effect of lighter quarks in the gluon fusion loop~\cite{Bishara:2016jga}.
The cross section for Higgs boson production in association with top quarks is taken to scale quadratically with $\kappat$.
The other production processes are taken to be independent of these couplings.
The coupling modifiers are described in the context of the \kappaframework~\cite{LHCHXSWG:YR3}:
\begin{linenomath*}
\begin{equation}
\kappa_{i} = \frac{y_{i}}{y_{i}^{\text{SM}}},
\end{equation}
\end{linenomath*}
where $y_i$ is the Higgs boson coupling to particle $i$.
The SM value of any $\kappa_i$ is equal to 1.

Recent developments in $\pt$ resummation procedures have allowed more accurate calculations of the $\pth$ spectrum when including the effects of lighter quarks on Higgs boson production via gluon fusion~\cite{Banfi:2013eda,Bozzi:2003jy,Becher:2010tm,Monni:2016ktx}.
The $\pth$ spectrum for gluon fusion has been calculated for simultaneous variations of $\kappac$ and $\kappab$~\cite{Bishara:2016jga}, taking into account the interference of the top quark loop with that from the bottom and charm quarks in the gluon fusion production loop, providing a novel approach to constrain these couplings via the $\pth$ spectrum.
We parameterize the variations computed in Ref.~\cite{Bishara:2016jga} with a quadratic polynomial for each bin of the $\pth$ spectrum.
The Higgs boson coupling to the top quark is fixed to its SM value in this model.
The calculations from Ref.~\cite{Bishara:2016jga} are given up to the scale of the Higgs boson mass, and thus the $\hbb$ channel (for which the lower limit of the $\pth$ spectrum is $350$\GeV) is not used as input for the results obtained with this model.

A second model producing simultaneous variations of $\kappat$, $\cg$, and $\kappab$ by adding dimension-6 operators to the SM Lagrangian has been built in Refs.~\cite{Grazzini:2017szg,Grazzini:2016paz}.
This study employs an analytic resummation performed up to next-to-next-to-leading-logarithmic (NNLL) order
in order to obtain the $\pth$ spectrum
at next-to-next-to-leading order+\allowbreak NNLL (NNLO+NNLL) accuracy.
The dimension-6 operator whose coefficient is $\cg$ yields a direct coupling of the Higgs field to the gluon field with the same underlying tensor structure as in the heavy-top mass limit.
In the SM, the value of $\cg$ equals 0.
The introduction of $\cg$ in the effective Lagrangian is given in Ref.~\cite{Grazzini:2016paz} and the inclusive cross section is given by $\sigma \simeq \left| 12\cg + \kappat \right|^2 \sigma^\text{SM}$.
Two other operators are included in the Lagrangian to describe modifications of the top and bottom Yukawa couplings with coefficients $\kappat$ and $\kappab$, respectively.
While the model allows simultaneous variation of all three coupling modifiers, we consider only simultaneous variations of $\kappat$ and $\cg$, and of $\kappat$ and $\kappab$.
The precomputed spectra from Ref.~\cite{Grazzini:2017szg} are used as input and parametrized using a quadratic polynomial.

\section{The CMS detector}

The central feature of the CMS apparatus is a superconducting solenoid of 6\unit{m} internal diameter, providing a magnetic field of 3.8\unit{T}. Within the solenoid volume are a silicon pixel and strip tracker, a lead tungstate crystal electromagnetic calorimeter, and a brass and scintillator hadron calorimeter, each composed of a barrel and two endcap sections. Forward calorimeters extend the pseudorapidity ($\eta$) coverage provided by the barrel and endcap detectors. Muons are detected in gas-ionization chambers embedded in the steel flux-return yoke outside the solenoid.
A more detailed description of the CMS detector, together with a definition of the coordinate system used and the relevant kinematic variables, can be found in Ref.~\cite{Chatrchyan:2008zzk}.

\section{Inputs to the combined analysis}

For all the analyses used as input to the combination ($\hgg$~\cite{Sirunyan:2018kta}, $\hzztofourl$~\cite{CMS_AN_2016-442}, and $\hbb$~\cite{CMS_AN_2016-366}), the data set corresponds to an integrated luminosity of $35.9\fbinv$ recorded by the CMS experiment in 2016.
The $\hbb$ decay channel is only included in the combination of the $\pth$ spectra, improving the measurements at the higher end of the distribution where the data from the $\hgg$ and $\hzz$ decay channels are limited.
All analyses provide the parametrization of the folding matrix $M_{ji}^{k}$ (which is the probability for an event in generator-level bin $i$ to be reconstructed in bin $j$ and category $k$) in terms of a common generator-level binning, that is used for the combined spectra.
Given the limited statistical precision in the individual channels, the results of the $\hzz$ and $\hbb$ channels individually are reported for a coarser binning, which is provided in Tables~\ref{tab:binningpth}--\ref{tab:binningptjet} for each of the observables.
This binning coincides with the binning at the reconstruction level.

The SM prediction for the differential cross sections is simulated with $\MGvATNLO$ v2.2.2~\cite{Alwall:2014hca} for each of the four dominant Higgs boson production modes: gluon-gluon fusion (\ggh), vector boson fusion, associated production with a $\PW$/$\zboson$ boson, and associated production with a top quark-antiquark pair.
A contribution from Higgs boson production in association with bottom quarks is not simulated, but included assuming its acceptance is equal to that from Higgs boson production via gluon fusion.
The matrix element calculation includes the emission of up to two additional partons and is performed at NLO accuracy in perturbative quantum chromodynamics (QCD).
Events are interfaced to \PYTHIA8.205~\cite{Sjostrand:2014zea} for parton showering and hadronization with the CUETP8M1~\cite{Skands:1695787} underlying event tune.
The matrix element calculation is matched to the parton shower following the prescription in Ref.~\cite{Frederix:2012ps}.
A weight depending on $\pth$ and $\njets$ is applied to simulated $\ggh$ events to match the predictions from the {\textsc{nnlops}} program~\cite{Hamilton:2012np, Kardos:2014dua}, as discussed in Ref.~\cite{Sirunyan:2018koj}.
The set of parton distribution functions used in all simulations is NNPDF3.0~\cite{Ball:2014uwa}.
The hadronic jets are clustered from the particle-flow candidates~\cite{Sirunyan:2017ulk} in the case of data and simulation, and from stable particles excluding neutrinos in the case of generated events, using the anti-$\kt$ clustering algorithm~\cite{Cacciari:2008gp} with a distance parameter of $0.4$.
The measurements are reported in terms of kinematic observables defined before the decay of the Higgs boson, \ie at the generator level.

Each of the analyses used as input to the combination corresponds to a different fiducial phase space definition and applies a different event categorization.
In the case of the $\hgg$ analysis, the fiducial phase space is defined by requiring the ratio of the leading (subleading) photon $\pt$ to the diphoton mass to be greater than $1/3$ ($1/4$).
In addition, for each photon candidate the scalar sum of the generator-level $\pt$ of stable particles contained in a cone of radius $\Delta R=0.3$ around the candidate is required to be less than 10\GeV, where $\Delta R = \sqrt{\smash[b]{(\Delta\eta)^2+(\Delta\phi)^2}}$ is the angular separation between particles and $\Delta\phi$ is the azimuthal angle between two particles in radians.
The selected photon pairs are categorized according to their estimated relative invariant mass resolution~\cite{Sirunyan:2018kta}.
In the case of the $\hzz$ analysis, the 4-lepton mass is required to be greater than 70\GeV, the leading $\zboson$ boson candidate invariant mass must be greater than 40\GeV, and leptons must be separated in angular space by at least $\Delta R > 0.02$.
Furthermore, at least two leptons must each have a $\pt>10$\GeV and at least one a $\pt > 20$\GeV.
The selected events are categorized according to their lepton configuration in the final state (4 electrons, 4 muons, or 2 electrons and 2 muons).
In the case of the $\hbb$ analysis, the analysis strategy requires the presence of a single anti-$\kt$ jet with a distance parameter of $0.8$, $\pt>450\,$GeV, and $\abs{\eta}<2.5$.
For this analysis, the data is not unfolded to a fiducial phase space.
Soft and wide-angle radiation is removed using the soft-drop grooming algorithm~\cite{Dasgupta:2013ihk,Larkoski:2014wba}.
The jet mass after application of the soft-drop algorithm, $\msd$, peaks close to the Higgs boson mass in the case of signal events.
To avoid finite-cone effects and the nonperturbative regime of the $\msd$ calculation, events are selected based on the dimensionless mass scale variable for QCD jets defined as $\rho=\log\left(\msd^2/\pt^2\right)$~\cite{Dasgupta:2013ihk}, which relates the jet $\pt$ to the jet mass.
Events with isolated electrons, muons, or \taulepton leptons with $\pt>10$\GeV and $\abs{\eta}<2.5$ are vetoed in order to reduce the background from SM electroweak processes, and events with a missing transverse momentum greater than $140$\GeV are vetoed in order to reduce the background from top quark-antiquark pair production.
Additionally, a selection criterion is applied based on the compatibility of the single anti-$\kt$ jet with having a two-prong substructure~\cite{Dolen:2016kst,Moult:2016cvt,Larkoski:2013eya,Thaler:2010tr}.
Events are categorized according to their likelihood of consisting of two $\bquark$ quarks, which is computed using the double-$\bquark$ tagger algorithm~\cite{Sirunyan:2017ezt}.

Minor modifications are applied to the individual analyses in Refs.~\cite{Sirunyan:2018kta,CMS_AN_2016-442,CMS_AN_2016-366} to provide the inputs used for the combination of differential observables.
For $\hgg$, an additional bin, $\pth>600$\GeV, is included in the $\pth$ spectrum.
For $\hzz$, the binning is modified for multiple kinematic observables to align with the binning of the $\hgg$ analysis.
Furthermore, the branching fractions of the two $\zboson$ bosons to the various lepton configurations are fixed to their SM values, whereas in Ref.~\cite{CMS_AN_2016-442} these are allowed to float.
For $\hbb$ the signal is split into two $\pt$ bins at the generator level:
the first with $350\le\pt<600$\GeV, where the lower limit has been extended downwards with respect to the individual analysis, and the second an overflow bin with $\pt\ge600$\GeV, which aligns with the binning of the other channels.
At the reconstruction level two bins are employed, with $450\le\pt<600$ and $\pt\ge600$\GeV, which is a slight modification with respect to the binning used in Ref.~\cite{CMS_AN_2016-366}.
The redefinition of the reconstructed $\pt$ categories necessitates a reevaluation of the background model, which is performed using the same procedure as in the original analysis.
For the purpose of the combination in this analysis, the fiducial measurements from the $\hgg$ and $\hzz$ channels are extrapolated to the inclusive phase space~\cite{Alwall:2014hca,Hamilton:2012np,Kardos:2014dua}.

\begin{table*}[htb!]
    \centering
    \topcaption{
        The reconstruction-level binning for $\pth$ for the $\hgg$, $\hzz$, and $\hbb$ channels.
        This binning coincides with the binning of the unfolded cross sections in which the individual results are reported.
        }
    \label{tab:binningpth}
    \cmsTable{
    \setlength{\tabcolsep}{5pt}
    \begin{tabular}{lccccccccc}
    Channel & \multicolumn{9}{l}{$\pth$ binning (\GeVns{})} \\[\cmsTabSkip]
    \hline
    $\hgg$
        & [0, 15)    & [15, 30)   & [30, 45)   & [45, 80)        & [80, 120)
        & [120, 200) & [200, 350) & [350, 600) & [600, $\infty$)
        \\
    $\hzz$
        & [0, 15) & [15, 30)
        & \multicolumn{2}{l}{[30,  80)}
        & \multicolumn{2}{l}{[80,  200)}
        & \multicolumn{3}{l}{[200, $\infty$)}
        \\
    $\hbb$
        & \multicolumn{7}{@{{}}c@{{}}}{None} & [350, 600) & [600, $\infty$)
        \\
    \end{tabular}
    }
    \end{table*}

\begin{table}[htb!]
    \centering
    \topcaption{
        The binning for $\njets$ for the $\hgg$ and the $\hzz$ channels.
        This binning coincides with the binning of the unfolded cross sections in which the individual results are reported.
        }
    \label{tab:binningnjets}
    \begin{tabular}{lccccc}
    Channel & \multicolumn{5}{l}{$\njets$ binning} \\[\cmsTabSkip]
    \hline
    $\hgg$ & 0 & 1 & 2 & 3 & $\ge$4 \\
    $\hzz$ & 0 & 1 & 2 & \multicolumn{2}{l}{$\ge$3} \\
    \end{tabular}
    \end{table}

\begin{table*}[htb!]
    \centering
    \topcaption{
        The binning for $\absy$ for the $\hgg$ and the $\hzz$ channels.
        This binning coincides with the binning of the unfolded cross sections in which the individual results are reported.
        }
    \label{tab:binningabsy}
    \begin{tabular}{lcccccc}
    Channel & \multicolumn{6}{l}{$\absy$ binning} \\[\cmsTabSkip]
    \hline
    $\hgg$ & [0.0, 0.15) & [0.15, 0.30) & [0.30, 0.60) & [0.60, 0.90) & [0.90, 1.20) & [1.20, 2.50] \\
    $\hzz$ & [0.0, 0.15) & [0.15, 0.30) & [0.30, 0.60) & [0.60, 0.90) & [0.90, 1.20) & [1.20, 2.50] \\
    \end{tabular}
    \end{table*}

\begin{table*}[htb!]
    \centering
    \topcaption{
        The binning for $\ptjet$ for the $\hgg$ and the $\hzz$ channels.
        This binning coincides with the binning of the unfolded cross sections in which the individual results are reported.
        }
    \label{tab:binningptjet}
    \begin{tabular}{lcccccc}
    Channel & \multicolumn{6}{l}{$\ptjet$ binning (\GeVns{})} \\[\cmsTabSkip]
    \hline
    $\hgg$ & [0, 30) & [30, 55) & [55, 95) & [95, 120) & [120, 200) & [200, $\infty$) \\
    $\hzz$ & [0, 30) & [30, 55) & [55, 95) & \multicolumn{3}{l}{ [95, $\infty$) } \\
    \end{tabular}
    \end{table*}

\section{Statistical analysis}
\label{sec:statisticalanalysis}

The cross sections are extracted through a simultaneous extended maximum likelihood fit to the diphoton mass, four-lepton mass, and $\msd$ distributions in all the analysis categories of the $\hgg$, $\hzz$, and $\hbb$ channels, respectively.

The number of expected signal events $n^\text{sig}$ in a given reconstructed kinematic bin $i$, given analysis category $k$ and given decay channel $m$ is obtained from:
\begin{linenomath*}
\begin{equation}
n_i^{\text{sig},\,km}(\vec{\Delta\sigma} | \vec{\theta})
= \sum_{j=1}^{n_\text{bins}^\text{gen}}
    \Delta\sigma_j \, L(\vec{\theta})
     \, \mathcal{B}^{m}
     \, M_{ji}^{km}(\vec{\theta}),
\label{eq:nsig}
\end{equation}
\end{linenomath*}
where:
\begin{itemize}
\item $j$ is a kinematic bin index at the generator level;
\item $n_\text{bins}^\text{gen}$ is the number of kinematic bins at the generator level, which is the same for all decay channels;
\item $\vec{\Delta\sigma}$ is the set of differential cross sections at the generator level, and $L$ is the integrated luminosity of the samples used in this analysis;
\item $\mathcal{B}^m$ is the branching fraction of the decay channel $m$. The overall effect of the branching fraction uncertainties on the combined spectra is below 1\%, and has been neglected.
\item $M_{ji}^{km}$ is the folding matrix, which is determined from Monte Carlo simulation;
note that the corresponding matrix $\vec{M}^{\,km}$ need not be square; the number of reconstructed bins may be smaller than the number of bins at the generator level; and
\item $\vec{\theta}$ is the set of nuisance parameters.
\end{itemize}
The bin-to-bin migrations are taken into account via the folding matrix, effectively allowing unfolding of the detector effects.
Following the prescription in Ref.~\cite{Hansen:LShape}, we find that no regularization of the unfolding procedure is needed.

An extended likelihood function for a single decay channel $m$ is constructed:
\begin{linenomath*}
\begin{multline}
\mathcal{L}_m
            (\vec{\Delta\sigma} | \vec{\theta})
            =
            \prod_{i=1}^{n_\text{bins}^{\text{reco},\,m}}
            \prod_{k=1}^{n_\text{cat}^m}
            \prod_{l=1}^{n_\mathcal{O}^m}
            \left(
            \text{pdf}_i^{\,km}(\mathcal{O}_l^m | \vec{\Delta\sigma}, \vec{\theta})
            \right)^{ N_\text{obs}^{iklm} }
            \\
       \times
            \text{Poisson}\left(
            N_\text{obs}^{ikm}
            \, \left| \,
            n_i^{\text{sig},\,km}(\vec{\Delta\sigma} | \vec{\theta})
            + n^{\text{bkg},\,km}_i(\vec{\theta})
            \right)\right.,
\label{eq:L_per_decaychannel}
\end{multline}
\end{linenomath*}
where:
\begin{itemize}
\item $\mathcal{O}^m$ is the observable, \ie the diphoton mass, the four-lepton mass, or $\msd$ for the $\hgg$, $\hzz$, and $\hbb$ decay channels, respectively;
\item $n_\text{bins}^{\text{reco},\,m}$ is the number of reconstructed bins,
$n_\text{cat}^m$ is the number of categories for the decay channel (see the individual analyses~\cite{Sirunyan:2018kta, CMS_AN_2016-442, CMS_AN_2016-366} for more details),
and $n_\mathcal{O}^m$ is the number of bins for observable $\mathcal{O}$;
\item $N_\text{obs}^{iklm}$ is the number of observed events reconstructed in kinematic bin $i$, category $k$ and observable bin $l$, and $N_\text{obs}^{ikm}$ is the same but summed over all bins of the observable;
\item $n^{\text{bkg},\,km}_i$ is the number of expected background events; and
\item $\text{pdf}_i^{\,km}(\mathcal{O}^m_l | \vec{\Delta\sigma}, \vec{\theta})$ is the probability density function for the observable, based on the signal and background distributions of the observable which are determined via simulation.
\end{itemize}
In order to combine the decay channels, the likelihoods for the individual decay channels are multiplied:
\begin{linenomath*}
\begin{equation}
\label{eq:fulllikelihood}
\mathcal{L}(\vec{\Delta\sigma} | \vec{\theta})
= \prod_{m=1}^{n_c} \mathcal{L}_m(\vec{\Delta\sigma} | \vec{\theta})
    \,
    \text{pdf}(\vec{\theta}),
\end{equation}
\end{linenomath*}
where $n_c$ is the number of decay channels included in the combination, $\mathcal{L}_m$ is the likelihood formula from Eq.~(\ref{eq:L_per_decaychannel}) specific to the decay channel $m$, and $\text{pdf}(\vec{\theta})$ is the probability density function of the nuisance parameters.
For the individual analyses, the number of categories, invariant mass bins, and even the number of reconstructed bins may differ, although the number of bins at the generator level and their binning need to be aligned between decay channels.
Note that a single common set of differential cross sections and nuisance parameters is fitted to the data in all decay channels simultaneously.

The test statistic $q$, which is asymptotically distributed as a $\chi^2$, is defined as~\cite{Cowan:2010js,CMS-NOTE-2011-005}:
\begin{linenomath*}
\begin{equation}
q(\vec{\Delta\sigma}) = -2 \, \ln \left(
    \frac{
        \mathcal{L}
            \left(
            \vec{\Delta\sigma} \left| \hat{\vec{\theta}}_{\vec{\Delta\sigma}}
            \right)\right.
        }{
        \mathcal{L}
            \left(
            \hat{\vec{\Delta\sigma}} \left| \hat{\vec{\theta}}
            \right)\right.
        }
\right).
\label{eq:TestStatisticQ}
\end{equation}
\end{linenomath*}
The quantities $\hat{\vec{\Delta\sigma}}$ and $\hat{\vec{\theta}}$ are the unconditional maximum likelihood estimates for the parameters $\vec{\Delta\sigma}$ and $\vec{\theta}$, respectively, while $\hat{\vec{\theta}}_{\vec{\Delta\sigma}}$ denotes the maximum likelihood estimate for $\vec{\theta}$ conditional on the values of $\vec{\Delta\sigma}$.

The Higgs boson coupling modifiers are fitted via a largely analogous procedure.
In the likelihood function~(\ref{eq:fulllikelihood}), the differential cross sections $\vec{\Delta\sigma}$ are replaced by parametrizations of theoretical spectra, instead of allowing them to be determined in the fit:
\begin{linenomath*}
\begin{equation}
    \vec{\Delta\sigma} \; \to \; \vec{\Delta\sigma}( \kappa_\text{a}, \kappa_\text{b} ),
\end{equation}
\end{linenomath*}
where $\kappa_\text{a}$ and $\kappa_\text{b}$ are the coupling modifiers to be fitted.

\section{Systematic uncertainties}
\label{sec:systematics}

The experimental systematic uncertainties from the input analyses are incorporated in the combination as nuisance parameters in the extended likelihood fit and are profiled.
Among the decay channels, correlations are taken into account for the systematic uncertainties in the jet energy scale and resolution, and the integrated luminosity.
Detailed descriptions of the experimental systematic uncertainties per decay channel can be found in Refs.~\cite{Sirunyan:2018kta,CMS_AN_2016-442,CMS_AN_2016-366}.

The measurement is made for the full phase space rather than limited to a fiducial phase space (as is the case for the original $\hgg$ and $\hzz$ analyses).
This means that the uncertainties in the acceptances for the individual analyses and in the branching fractions may affect the results.
The effect of the acceptance uncertainties per bin on the overall uncertainty, including the effect of the Higgs coupling modifiers on the acceptances, is less than 1\% and so this is neglected in the combination.
For certain measurements the production cross sections of non-$\ggh$ production modes are assumed to be their respective SM value.
In these cases, the uncertainty in the inclusive production cross section from non-$\ggh$ modes, determined to be about 2.1\%~\cite{deFlorian:2016spz}, has been taken into account as a nuisance parameter.

The theoretical predictions described in Section~\ref{sec:theory} are subject to theoretical uncertainties from the renormalisation scale $\muR$ and the factorisation scale $\muF$.
The standard approach to evaluate the impact of these uncertainties is to compute an envelope of scale variations, and to assign the extrema of the envelope as the uncertainty.
To this end, $\muR$ and $\muF$ are independently varied between $0.5$, $1$, and $2$ times their nominal value, whereas the fraction $\frac{\muR}{\muF}$ is constrained not to be less than $0.5$ or greater than $2.0$.
As the theoretical spectra in the $\kappat$/$\cg$/$\kappab$ case and the $\kappac$/$\kappab$ case contain a resummation, the uncertainty in the resummation scale $Q$ is also considered, and it is evaluated by varying $Q$ from $0.5$ to $2$ times its central value (while keeping $\muF$ and $\muR$ at their respective central values).
The theoretical uncertainties are assigned by applying the minimum and maximum scale variations per bin.
The resulting uncertainties for the spectra under variations of $\kappab$ and $\kappac$ and variations of $\kappat$, $\cg$, and $\kappab$ are shown in Tables~\ref{tab:TheoryUncertainties_kappab_kappac} and \ref{tab:TheoryUncertainties_kappat_kappag}, respectively.

\begin{table*}[htb!]
\centering
\topcaption{
    Uncertainties in the predicted $\pth$ spectra related to variations of theory parameters for the $\kappab$ and $\kappac$ case.
    }
\label{tab:TheoryUncertainties_kappab_kappac}
\begin{tabular}{lccccc}
Binning (\GeVns{}) & [0, 15) & [15, 30) & [30, 45) & [45, 80) & [80, 120) \\
\hline
$\Delta^\text{scale}$ (\%) & 8.9\% & 6.6\% & 18.1\% & 22.0\% & 21.6\% \\
\end{tabular}
\end{table*}

\begin{table*}[htb!]
\centering
\topcaption{
    Uncertainties in the predicted $\pth$ spectra related to variations of theory parameters for the $\kappat$, $\cg$, and $\kappab$ case.
    }
\label{tab:TheoryUncertainties_kappat_kappag}
\setlength{\tabcolsep}{3pt}
\cmsTable{
    \begin{tabular}{lccccccccc}
    Binning (\GeVns{}) & [0, 15) & [15, 30) & [30, 45) & [45, 80) & [80, 120) & [120, 200) & [200, 350) & [350, 600) & [600, 800) \\
    \hline
    $\Delta^\text{scale}$ (\%) & 12.7\% & 7.4\% & 9.5\% & 12.8\% & 17.4\% & 19.3\% & 20.9\% & 23.4\% & 8.2\% \\
    \end{tabular}
    }
\end{table*}

Theoretical uncertainties are subject to bin-to-bin correlations.
We adopt a procedure that produces a correlation coefficient $\rho_{ab}$ directly from the individual scale variations:
\begin{linenomath*}
\begin{equation}
\rho_{ab} =
\frac{
    \sum_i ( \sigma_{a, i} - \overline{\sigma}_a ) ( \sigma_{b, i} - \overline{\sigma}_b )
    }{
    \sqrt{
        \sum_i ( \sigma_{a, i} - \overline{\sigma}_a )^2
        \sum_i ( \sigma_{b, i} - \overline{\sigma}_b )^2
        }
    },
\end{equation}
\end{linenomath*}
where $\sigma_{a (b), i}$ is the cross section in bin $a$ ($b$) of the $i$th scale variation, $\overline{\sigma}_{a (b)}$ is the mean cross section in bin $a$ ($b$), and $\rho_{ab}$ is the resulting correlation coefficient between bin $a$ and $b$.
The correlation structure is characterized by strong correlations among bins at moderate $\pth$ ($15 \leq \pth \leq 600\GeV$).
Only the bins with $\pth<15$ and $\pth>600$\GeV are anti-correlated with the bins at moderate $\pth$.

\section{Results}

\subsection{Total cross section and \texorpdfstring{$\BRgamgam/\BRZZ$}{BRgg/BRZZ}}
\label{sec:ratioOfBrsTotalXS}

\begin{figure}[hbt!]
  \begin{center}
    \includegraphics[width=\cmsFigWidth]{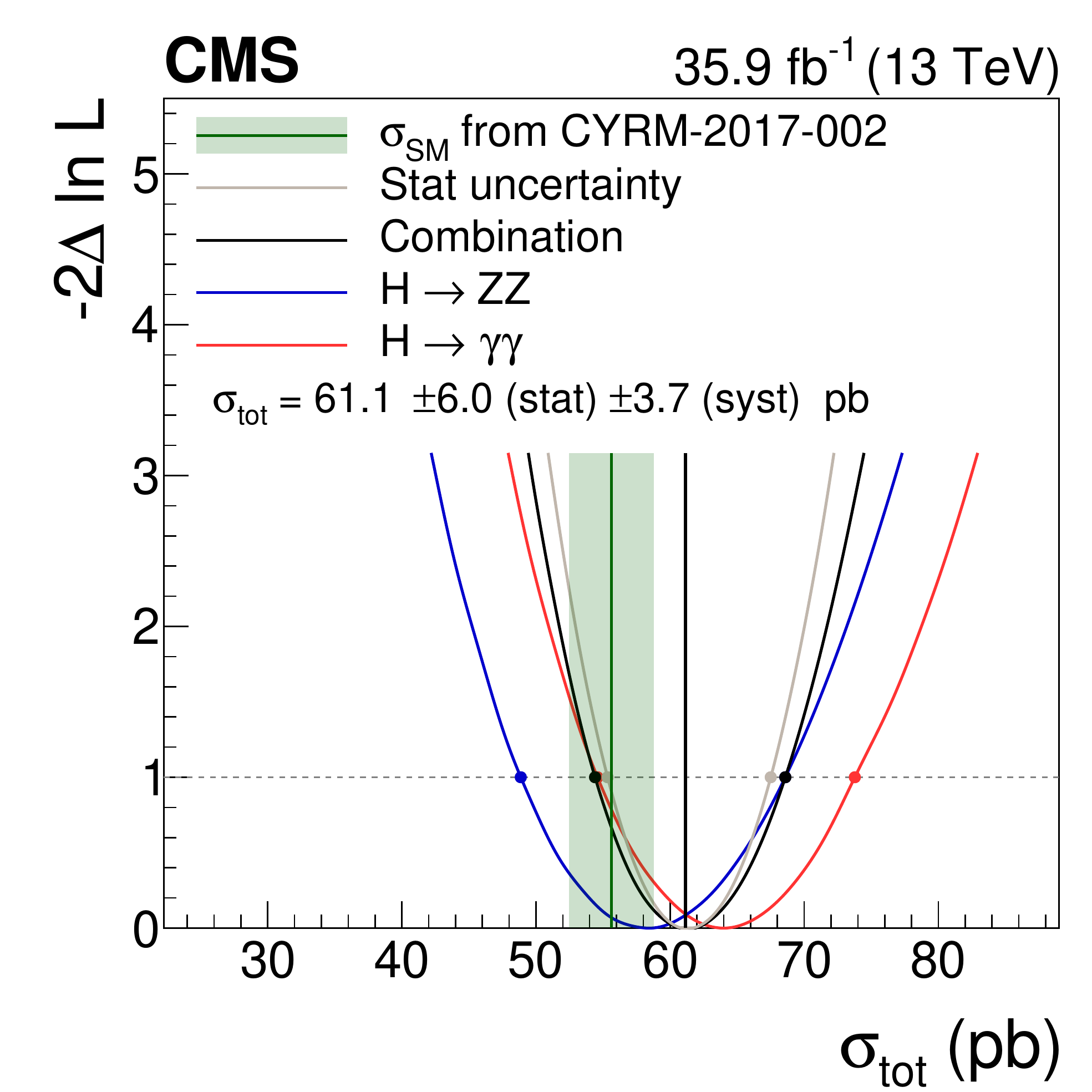}
    \includegraphics[width=\cmsFigWidth]{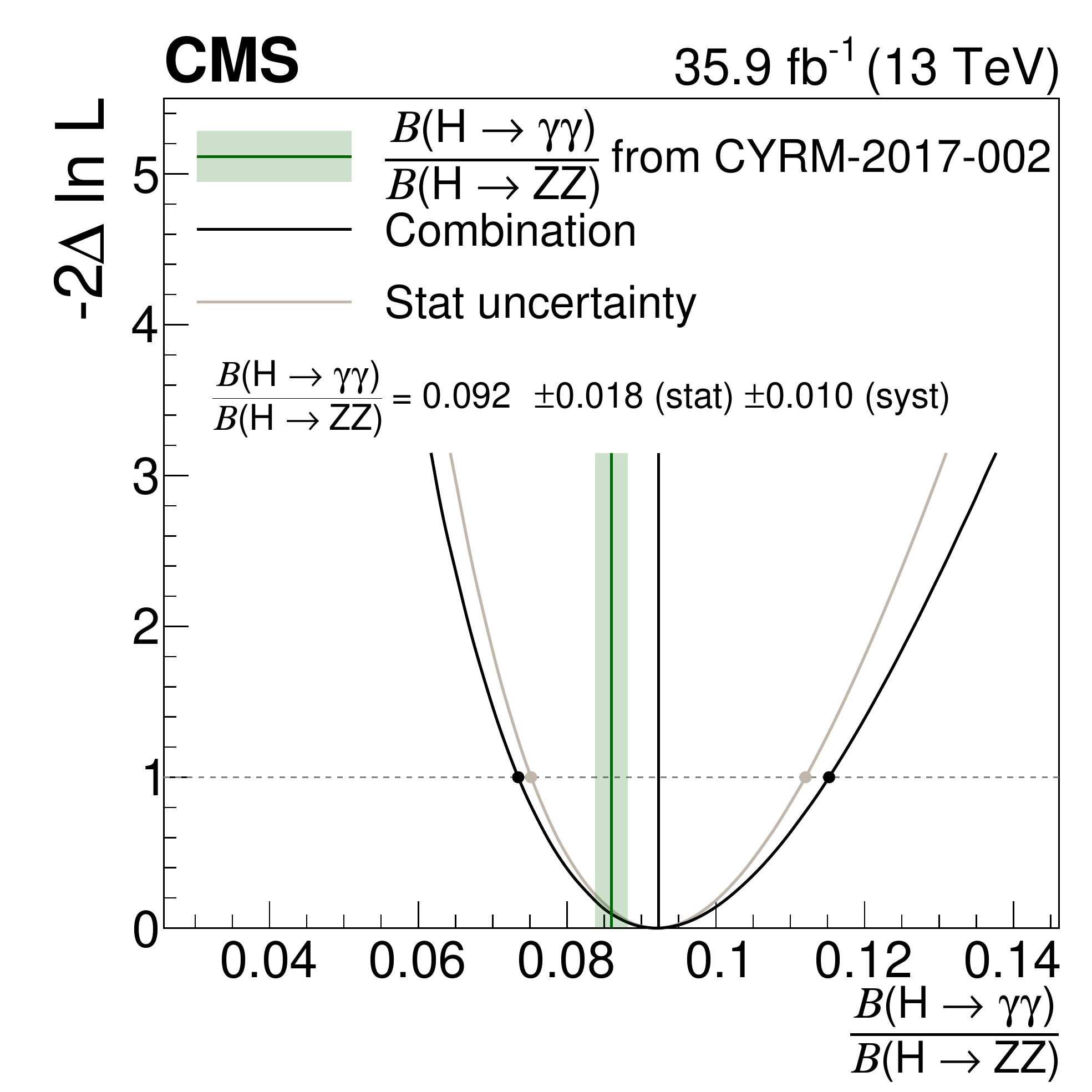}
    \caption{
        Scan of the total cross section $\sigma_\text{tot}$ (\cmsLeft) and of the ratio of branching fractions $\BRgamgam/\BRZZ$ (\cmsRight), based on a combination of the $\hgg$ and $\hzz$ analyses.
        The markers indicate the one standard deviation confidence interval.
        \textit{CYRM-2017-002} refers to Ref.~\cite{deFlorian:2016spz}.
        }
    \label{fig:RatioOfbrsAndTotalXSscan}
  \end{center}
\end{figure}

The total cross section for Higgs boson production, based on a combination of the $\hgg$ and $\hzz$ channels, is measured to be $61.1   \pm 6.0 \stat   \pm 3.7 \syst  $\pb, obtained by applying the treatment described in Section 4 to the inclusive cross section (\ie with a single bin, both at generator and at reconstruction level).
The measured total cross sections from the individual channels are $64.0\pm9.6$\pb for $\hgg$ and $58.2\pm9.8$\pb for $\hzz$; the combination improves the precision by 27\% with respect to the $\hgg$ channel individually.
The likelihood scans for the individual decay channels and their combination are shown in Fig.~\ref{fig:RatioOfbrsAndTotalXSscan}~(\cmsLeft).
The combination result agrees with the SM value of $55.6\pm2.5$\pb~\cite{deFlorian:2016spz}.

A measurement of the branching fraction for one decay channel is degenerate with a measurement of the total cross section.
However, the ratio of branching fractions for two decay channels can be measured while profiling the total cross section.
The ratio of the $\hgg$ and $\hzz$ branching fractions, $\BRgamgam/\BRZZ$, is measured to be
$0.092   \pm 0.018 \stat   \pm 0.010 \syst  $.
This is in agreement with the SM prediction of $0.086 \pm 0.002$~\cite{deFlorian:2016spz}.
The likelihood scan for $\BRgamgam/\BRZZ$ is shown in Fig.~\ref{fig:RatioOfbrsAndTotalXSscan}~(\cmsRight).

\subsection{Combinations of differential observables}
\label{sec:noncouplingresults}

\begin{figure}[htbp!]
  \begin{center}
    \includegraphics[width=\cmsFigWidth]{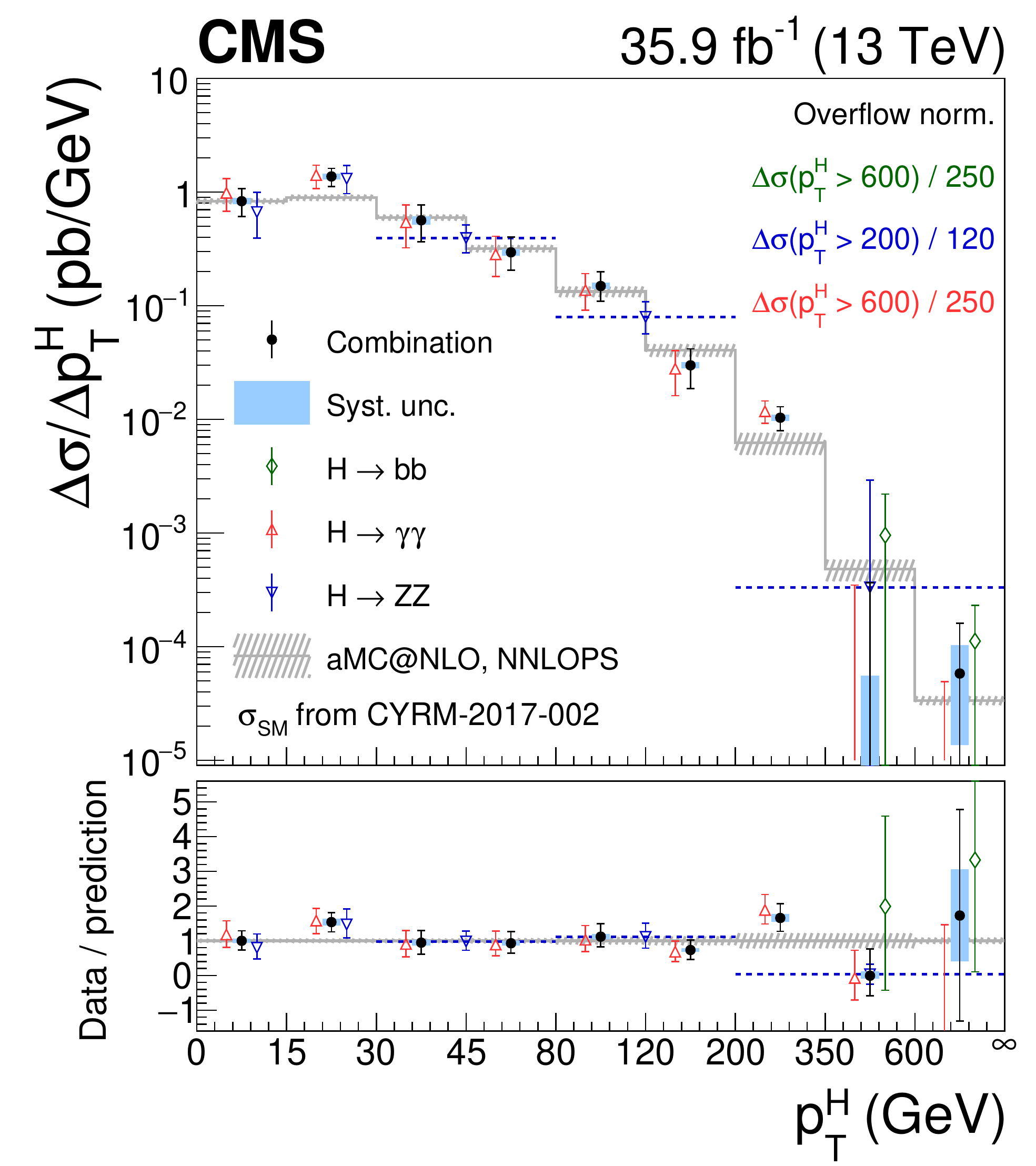}
    \includegraphics[width=\cmsFigWidth]{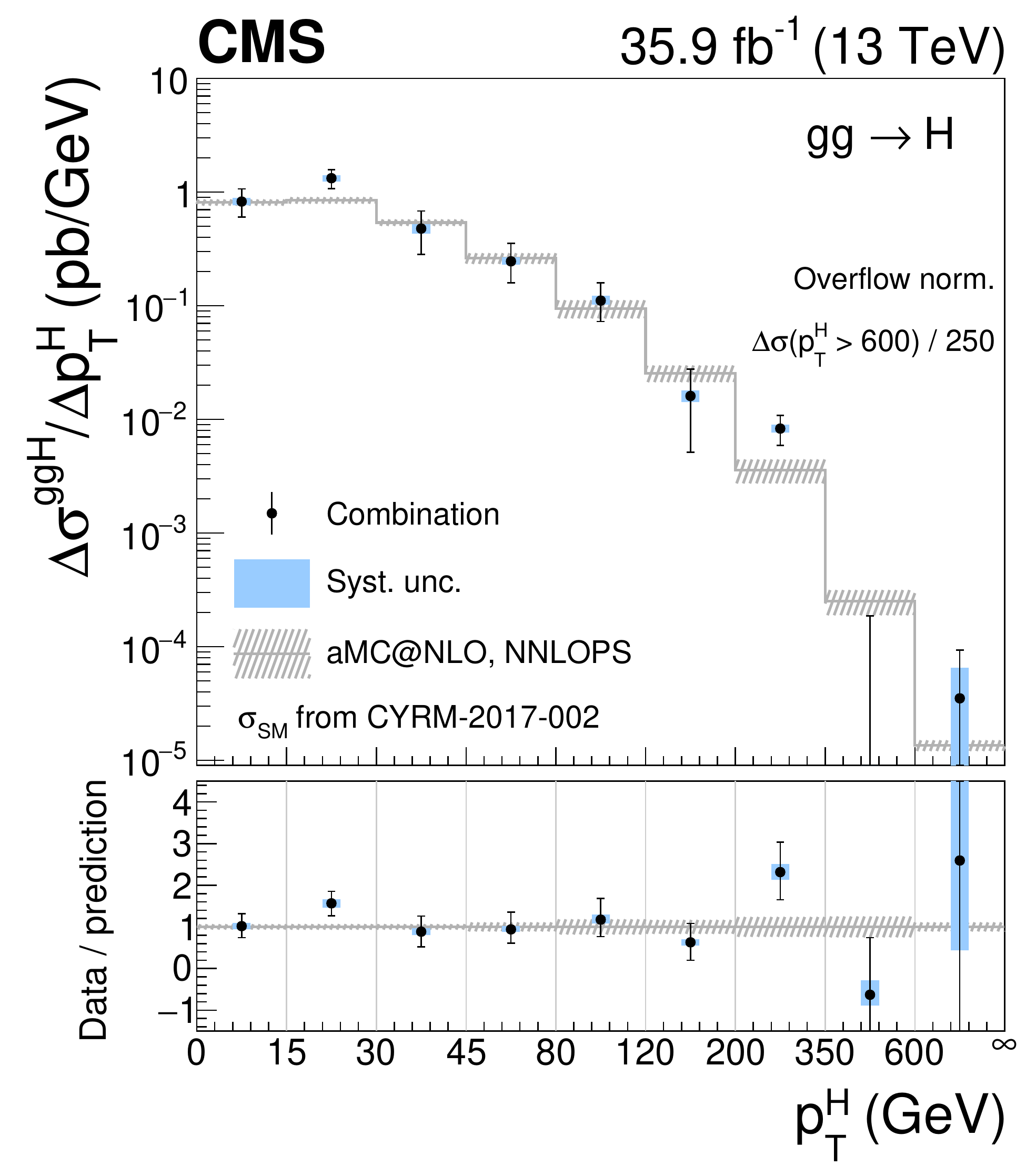}
    \caption{
        Measurement of the total differential cross section (\cmsLeft) and the differential cross section of gluon fusion (\cmsRight) as a function of $\pth$. The combined spectrum is shown as black points with error bars indicating a 1 standard deviation uncertainty. The systematic component of the uncertainty is shown by a blue band. The spectra for the $\hgg$, $\hzz$, and $\hbb$ channels are shown in red, blue, and green, respectively.
        The dotted horizontal lines in the $\hzz$ channel indicate the coarser binning of this measurement.
        The rightmost bins of the distributions are overflow bins; the normalizations of the cross sections in these bins are indicated in the figure.
        \textit{CYRM-2017-002} refers to Ref.~\cite{deFlorian:2016spz}.
        }
    \label{fig:CombinedSpectra_pth}
  \end{center}
\end{figure}

\begin{figure}[hbt!]
  \begin{center}
    \includegraphics[width=\cmsFigWidth]{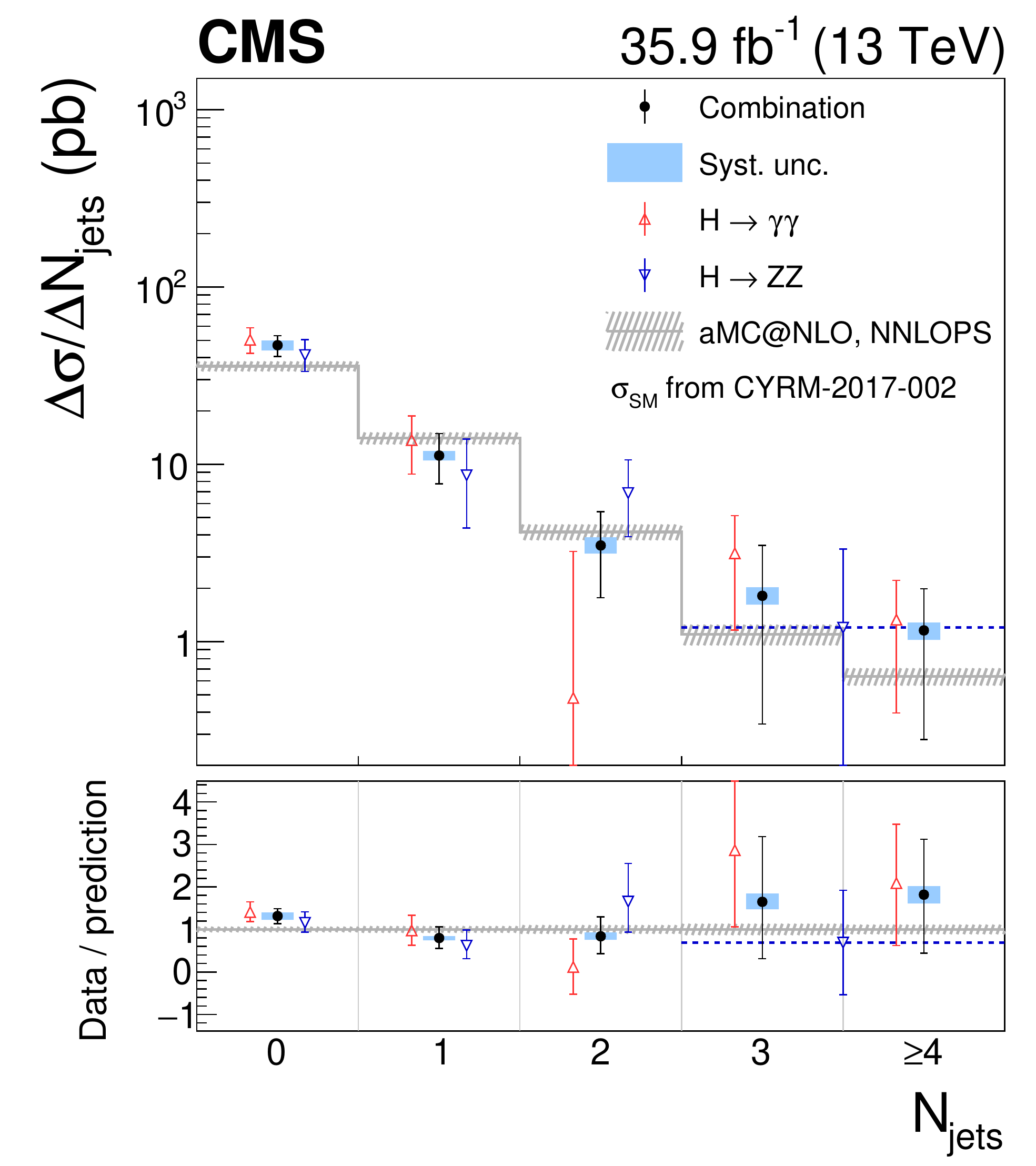}
    \caption{
        Measurement of the differential cross section as a function of $\njets$. The combined spectrum is shown as black points with error bars indicating a 1 standard deviation uncertainty. The systematic component of the uncertainty is shown by a blue band. The spectra for the $\hgg$ and $\hzz$ channels are shown in red and blue, respectively.
        The dotted horizontal lines in the $\hzz$ channel indicate the coarser binning of this measurement.
        \textit{CYRM-2017-002} refers to Ref.~\cite{deFlorian:2016spz}.
        }
    \label{fig:CombinedSpectra_njets}
  \end{center}
\end{figure}

\begin{figure}[hbt!]
  \begin{center}
    \includegraphics[width=\cmsFigWidth]{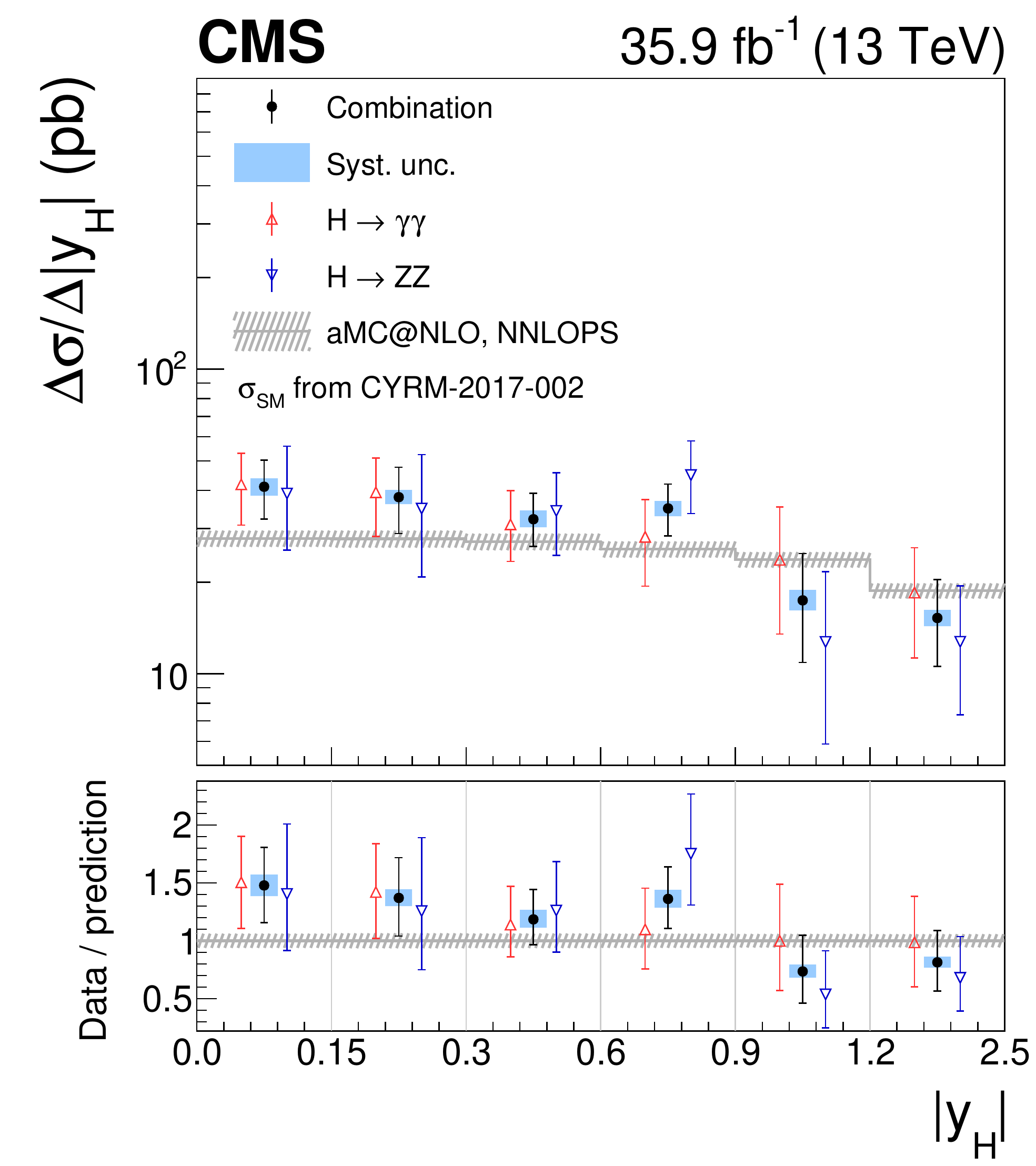}
    \caption{
        Measurement of the differential cross section as a function of $\absy$. The combined spectrum is shown as black points with error bars indicating a 1 standard deviation uncertainty. The systematic component of the uncertainty is shown by a blue band. The spectra for the $\hgg$ and $\hzz$ channels are shown in red and blue, respectively.
        \textit{CYRM-2017-002} refers to Ref.~\cite{deFlorian:2016spz}.
        }
    \label{fig:CombinedSpectra_rapidity}
  \end{center}
\end{figure}

\begin{figure}[hbt!]
  \begin{center}
    \includegraphics[width=\cmsFigWidth]{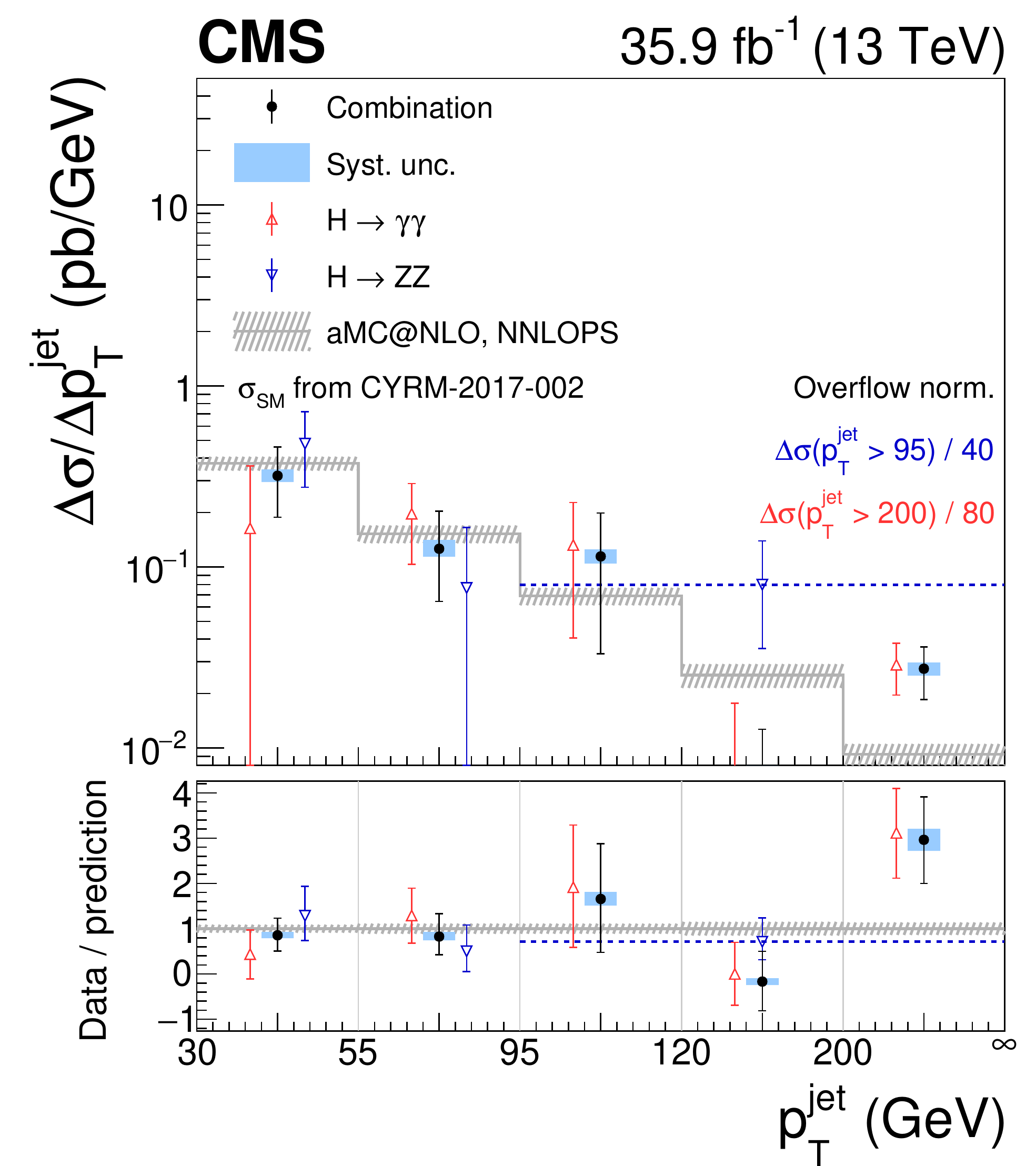}
    \caption{
        Measurement of the differential cross section as a function of $\ptjet$. The combined spectrum is shown as black points with error bars indicating a 1 standard deviation uncertainty. The systematic component of the uncertainty is shown by a blue band. The spectra for the $\hgg$ and $\hzz$ channels are shown in red and blue, respectively.
        The dotted horizontal lines in the $\hzz$ channel indicate the coarser binning of this measurement.
        The rightmost bin of the distribution is an overflow bin; the normalization of the cross section in that bin is indicated in the figure.
        \textit{CYRM-2017-002} refers to Ref.~\cite{deFlorian:2016spz}.
        }
    \label{fig:CombinedSpectra_ptjet}
  \end{center}
\end{figure}

The unfolded differential cross sections for the observables $\pth$, $\njets$, $\absy$, and $\ptjet$ are shown in
Figs.~\ref{fig:CombinedSpectra_pth}, \ref{fig:CombinedSpectra_njets}, \ref{fig:CombinedSpectra_rapidity}, and \ref{fig:CombinedSpectra_ptjet}, respectively.
Figure~\ref{fig:CombinedSpectra_pth} (\cmsRight) shows the differential cross section of $\pth$ for Higgs boson production via gluon fusion;
for this result, the non-gluon-fusion production modes are considered to be background, constrained to the SM predictions with their respective uncertainties.
The numerical values for the spectra in Figs.~\ref{fig:CombinedSpectra_pth}--\ref{fig:CombinedSpectra_ptjet} are given in \cmsAppend~\ref{sec:tables} and the corresponding bin-to-bin correlation matrices are given in \cmsAppend~\ref{sec:binToBinCorrelationMatrices}.
For the observables $\pth$, $\njets$, and $\ptjet$, the rightmost bin is an overflow bin, which is normalized by the bin width of the second-to-rightmost bin.
Overall no significant deviations from the SM predictions are observed.
For the $\pth$ spectrum, the dominant source of uncertainty is the statistical one; in particular, the systematic uncertainty is about half the statistical uncertainty in the rightmost bin, and much smaller than the statistical uncertainty in all other bins.
The total uncertainty in the combination per bin varies between 30 and 40\%.
Compared to the measurement in the $\hgg$ channel alone, the decrease in uncertainty achieved by the combination is most notable in the low-$\pt$ region.
The contribution of the $\hbb$ channel to the overall precision of the combination is most significant in the last $\pth$ bin.

\subsection{Fits of Higgs boson coupling modifiers: \texorpdfstring{$\kappab$}{kb} vs. \texorpdfstring{$\kappac$}{kc}}
\label{sec:ResultsKappabKappac}

\begin{figure}[hbt!]
  \begin{center}
    \includegraphics[width=\cmsFigWidth]{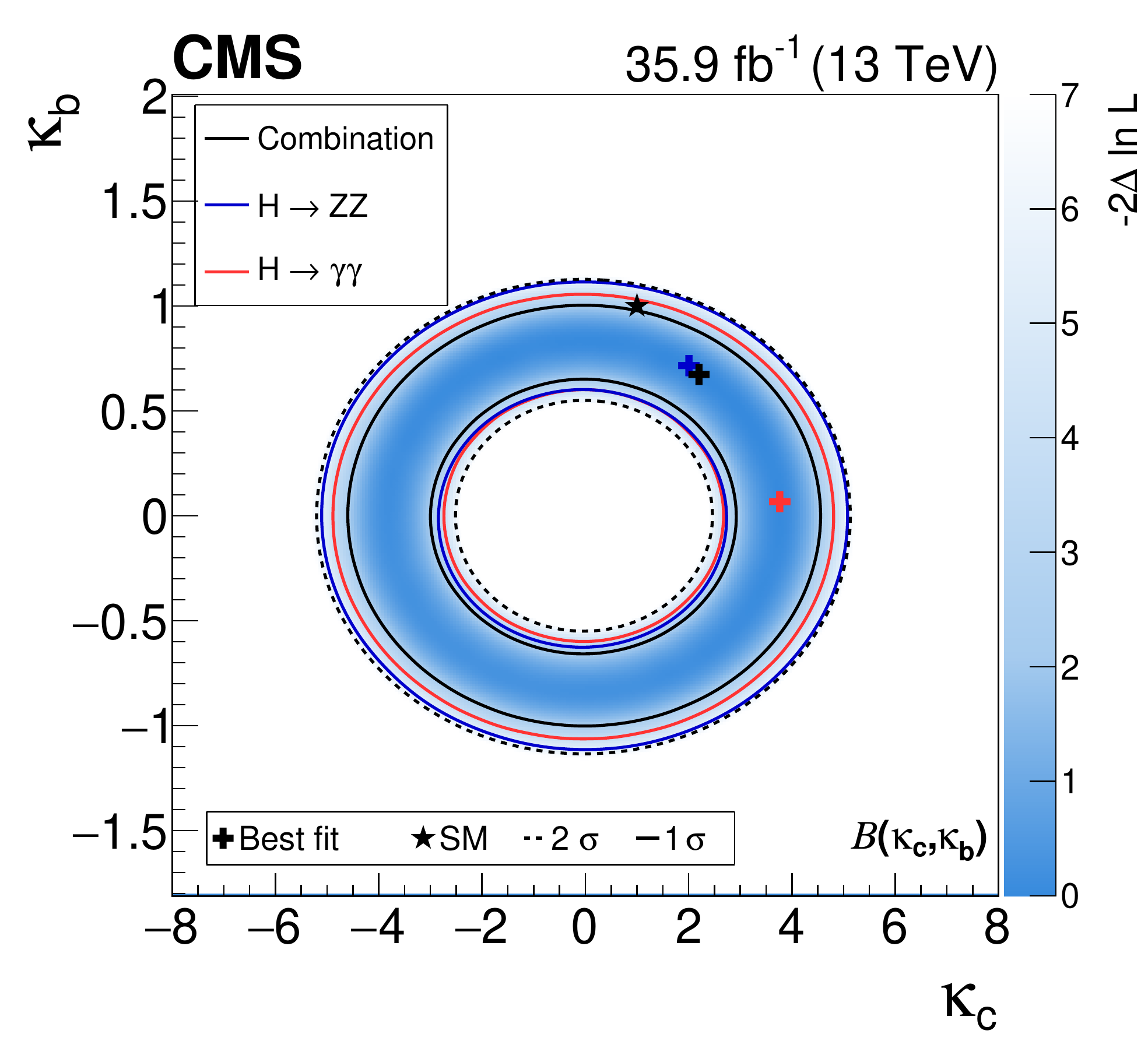}
    \includegraphics[width=\cmsFigWidth]{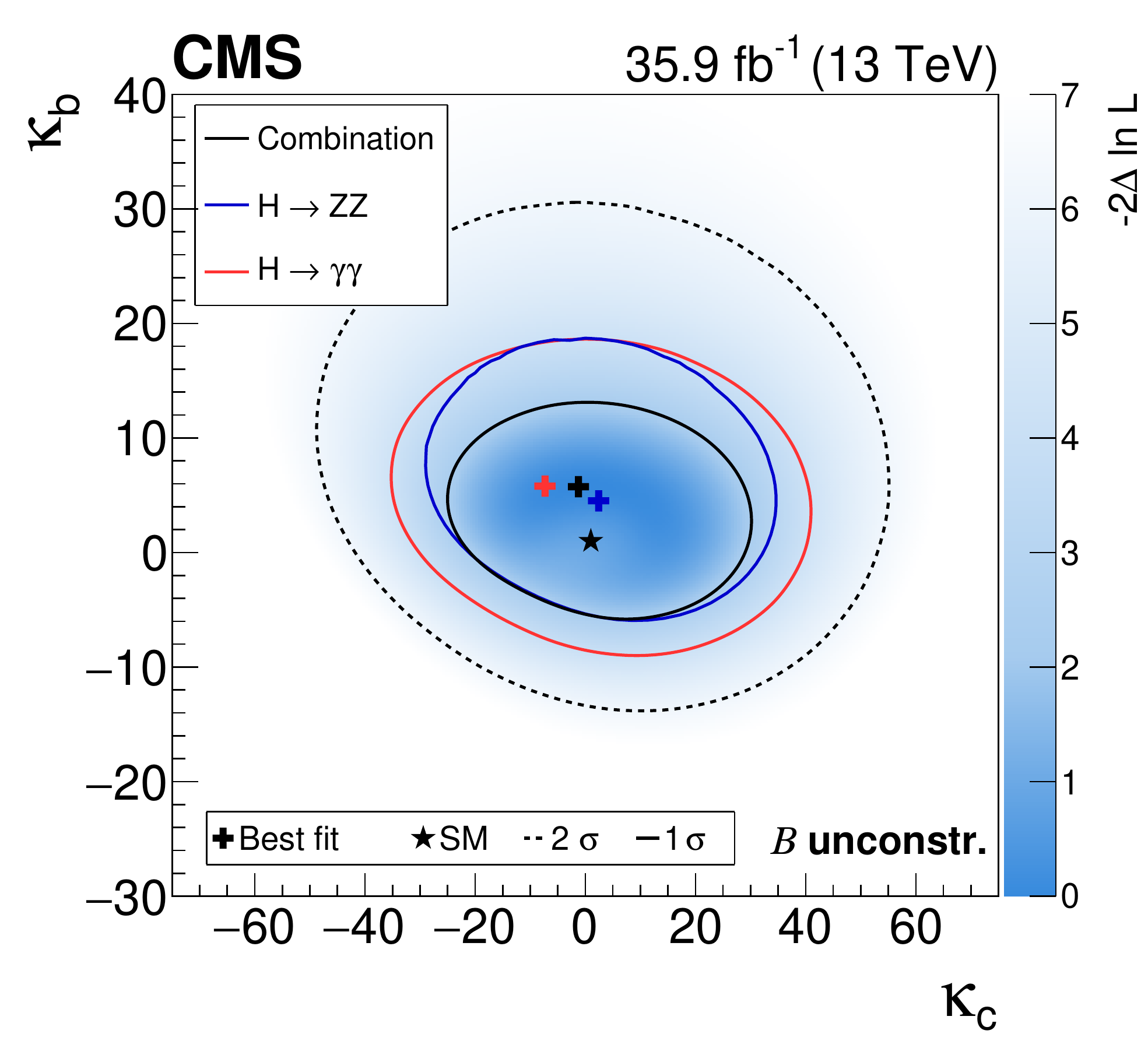}
        \caption{
        Simultaneous fit to data for $\kappab$ and $\kappac$, assuming a coupling dependence of the branching fractions (\cmsLeft) and the branching fractions implemented as nuisance parameters with no prior constraint (\cmsRight).
        The one standard deviation contour is drawn for the combination ($\hgg$ and $\hzz$), the $\hgg$ channel, and the $\hzz$ channel in black, red, and blue, respectively.
        For the combination the two standard deviation contour is drawn as a black dashed line, and the shading indicates the negative log-likelihood, with the scale shown on the right hand side of the plots.
        }
    \label{fig:scans_kappabkappac_nominal}
  \end{center}
\end{figure}

Figure~\ref{fig:scans_kappabkappac_nominal} (\cmsLeft) shows the one and two standard deviation contours of the fits of the $\kappab/\kappac$ parametrization from Ref.~\cite{Bishara:2016jga} to data, assuming the branching fractions are dependent on the Higgs boson couplings, \ie, $\mathcal{B} = \mathcal{B}(\kappab, \kappac)$, and that there are no beyond-the-SM contributions.
The substructure on the combined scan shows a ring shape around the origin, in agreement with the SM prediction within one standard deviation.

In order to assess the constraint obtained only from the knowledge of the $\pth$ distribution, the total width and the overall normalization are profiled in the fit.
This is effectively accomplished by implementing the branching fractions for the $\hgg$ and $\hzz$ channels as nuisance parameters with no prior constraint, \ie as free parameters.
The result of this fit is shown in Fig.~\ref{fig:scans_kappabkappac_nominal} (\cmsRight).
As expected, the range of allowed values of $\kappab$ and $\kappac$ is much wider than in the case of coupling-dependent branching fractions.

\begin{figure}[!hbt]
  \begin{center}
    \includegraphics[width=\cmsFigWidth]{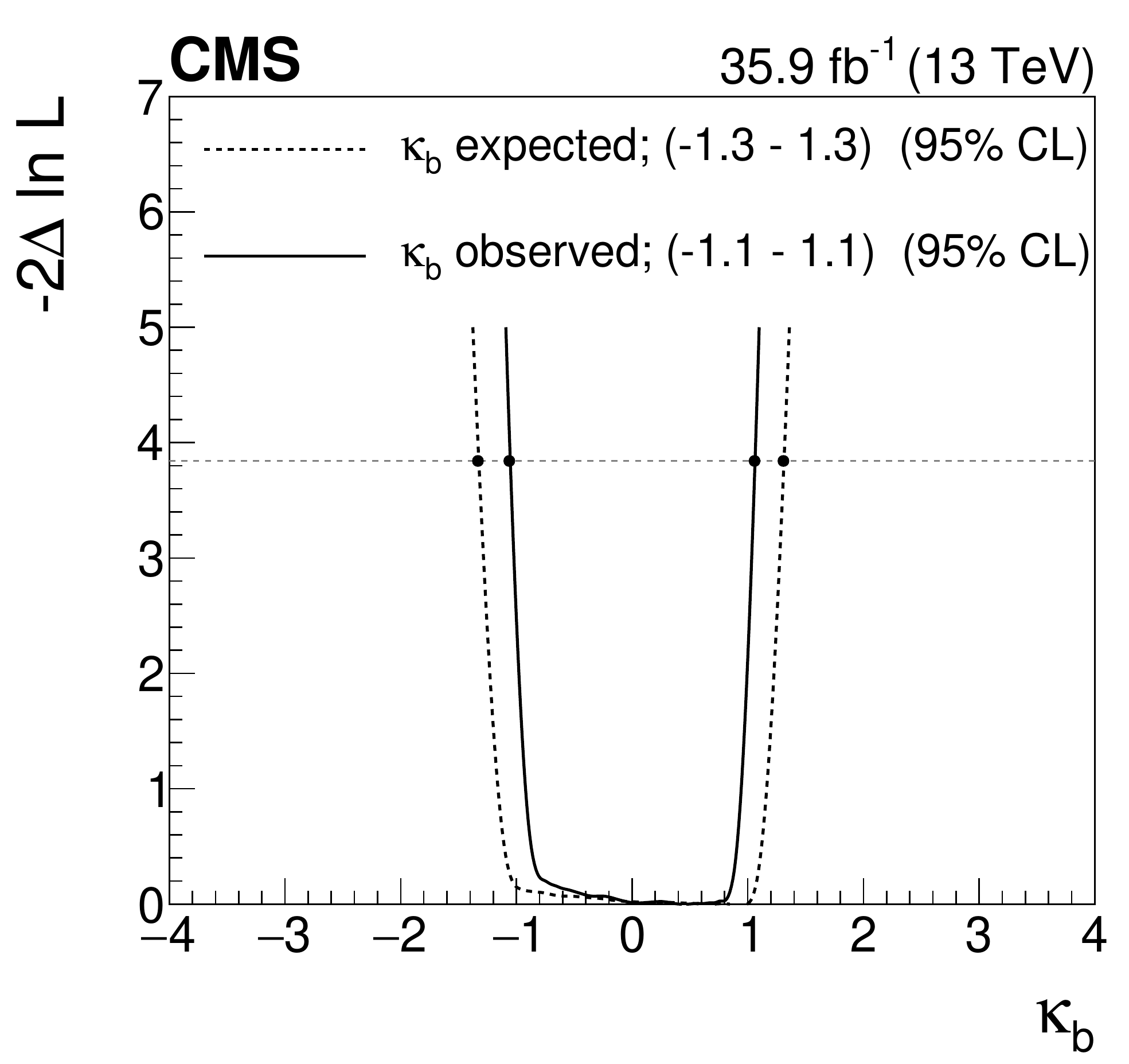}
    \includegraphics[width=\cmsFigWidth]{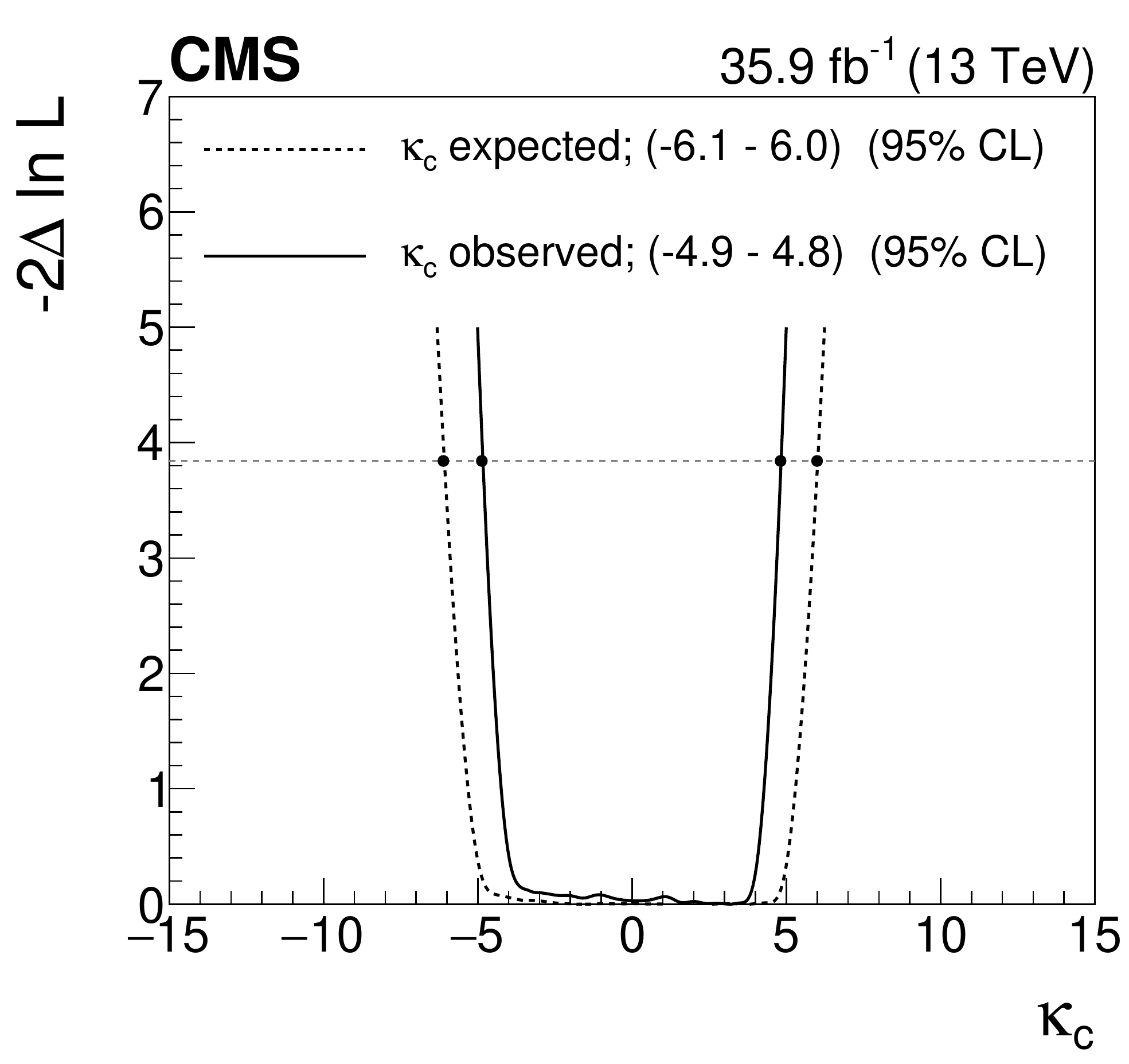}
    \caption{
        Likelihood scan of $\kappab$ while profiling $\kappac$ (\cmsLeft), and of $\kappac$ while profiling $\kappab$ (\cmsRight).
        The filled markers indicate the limits at 95\% \CL.
        The branching fractions are considered dependent on the values of the couplings.
        }
    \label{fig:scans_kappabkappac_oneDimScans}
  \end{center}
\end{figure}

\begin{figure}[!hbt]
  \begin{center}
    \includegraphics[width=\cmsFigWidth]{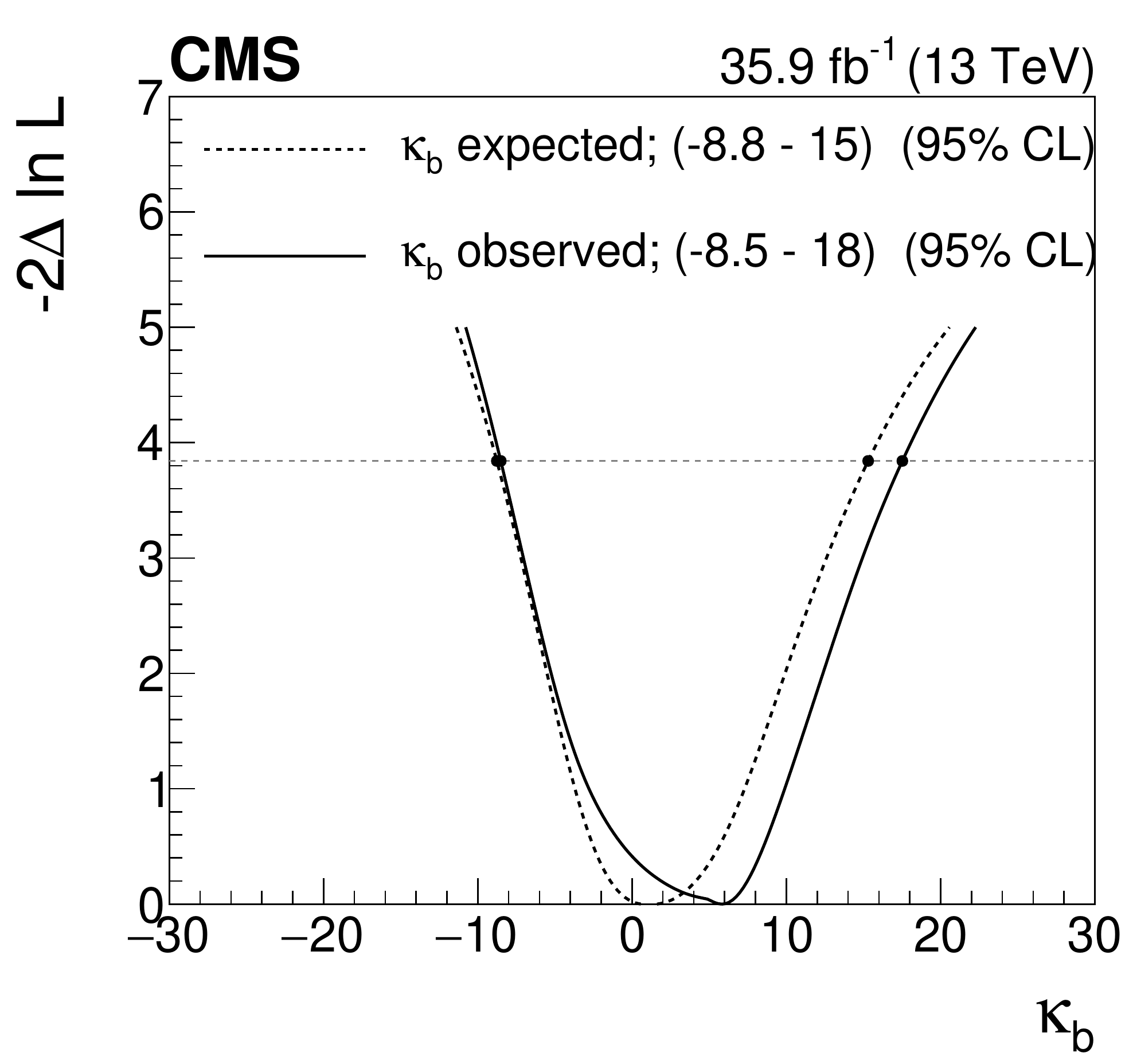}
    \includegraphics[width=\cmsFigWidth]{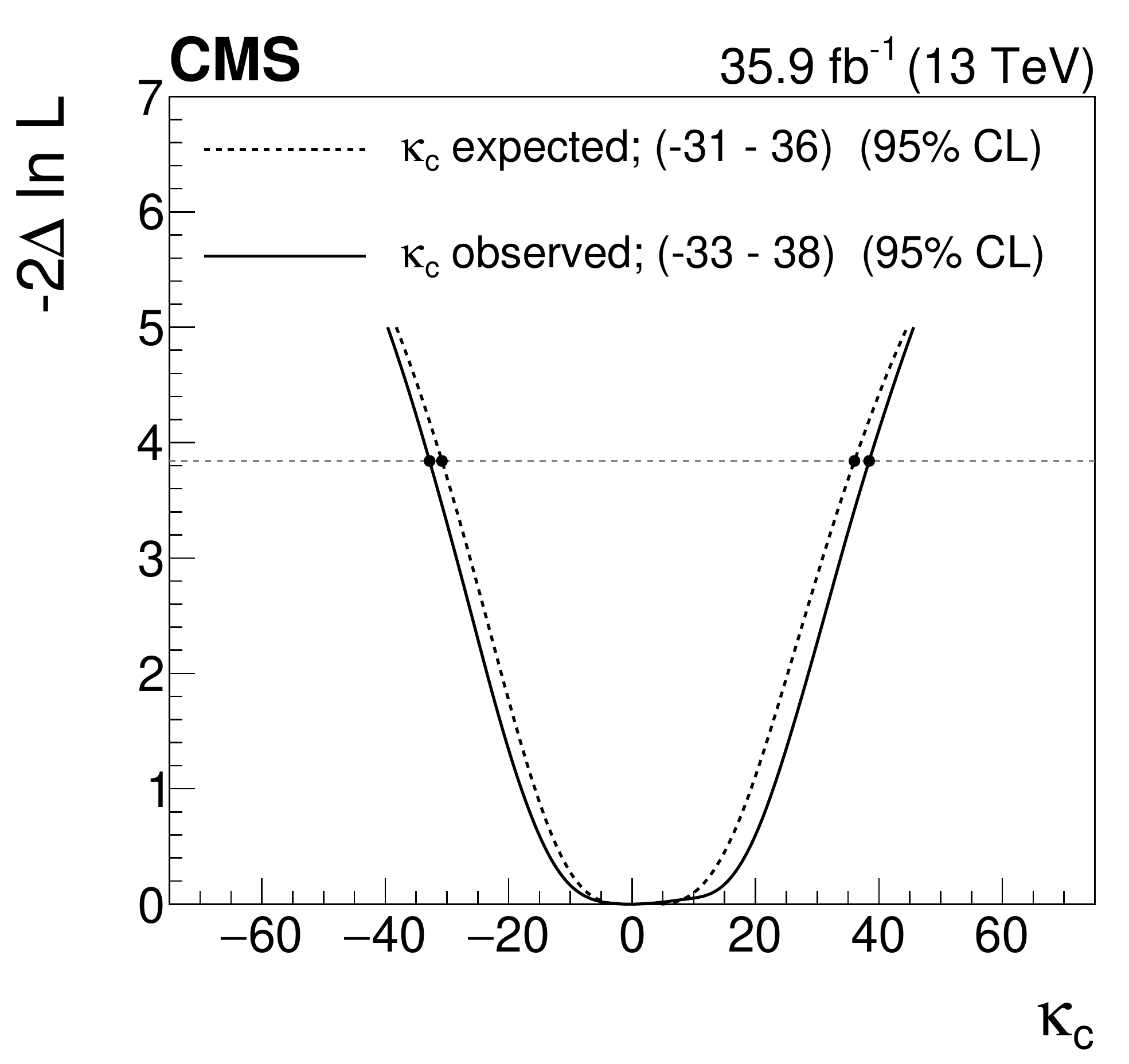}
    \caption{
        Likelihood scan of $\kappab$ while profiling $\kappac$ (\cmsLeft), and of $\kappac$ while profiling $\kappab$ (\cmsRight).
        The filled markers indicate the limits at 95\% \CL.
        The branching fractions are implemented as nuisance parameters with no prior constraint.
        }
    \label{fig:scans_kappabkappac_oneDimScans_scenario2}
  \end{center}
\end{figure}

Confidence intervals can be set on $\kappab$ and $\kappac$ by profiling one coupling and scanning over the other.
The results of these single-coupling scans are shown in Figs.~\ref{fig:scans_kappabkappac_oneDimScans} and \ref{fig:scans_kappabkappac_oneDimScans_scenario2}.
The observed (expected) limits at 95\% \CL in the one-dimensional scans are:
\begin{linenomath*}
\begin{equation}
\label{eq:kappasensitivity}
\begin{aligned}
\kappabLeftObserved < \kappab < \kappabRightObserved  &&(\kappabLeftAsimov < \kappab < \kappabRightAsimov ),
\\
\kappacLeftObserved < \kappac < \kappacRightObserved &&(\kappacLeftAsimov < \kappac < \kappacRightAsimov ),
\end{aligned}
\end{equation}
\end{linenomath*}
in the case of branching fractions that depend on $\kappab$ and $\kappac$, and
\begin{linenomath*}
\begin{equation}
\label{eq:kappasensitivity_floatingBRs}
\begin{aligned}
\kappabLeftObservedFLOATINGBRS < \kappab < \kappabRightObservedFLOATINGBRS  &&(\kappabLeftAsimovFLOATINGBRS < \kappab < \kappabRightAsimovFLOATINGBRS ),\\
\kappacLeftObservedFLOATINGBRS < \kappac < \kappacRightObservedFLOATINGBRS &&(\kappacLeftAsimovFLOATINGBRS < \kappac < \kappacRightAsimovFLOATINGBRS ),
\end{aligned}
\end{equation}
\end{linenomath*}
in the case of the branching fractions implemented as nuisance parameters with no prior constraint.
For the coupling-dependent branching fractions, the results are shaped predominantly by the constraints from the total width rather than by distortions of the $\pth$ spectrum.
If the branching fractions are fixed to their SM expectations, the one-dimensional scans yield the following expected limits at 95\% \CL:
\begin{linenomath*}
\begin{equation}
\label{eq:kappasensitivity_fixedSMBRs}
\begin{aligned}
-3.5 < &\kappab < 5.1,\\
-13 < &\kappac < 15.
\end{aligned}
\end{equation}
\end{linenomath*}
These intervals are comparable to those in Ref.~\cite{Bishara:2016jga}, where $\kappac \in [ -16, 18 ]$ at 95\% \CL, noting that the results here are based on a larger data set.
The intervals obtained are competitive with the intervals from other direct search channels summarized in Section~\ref{sec:introduction}.

\subsection{Fits of Higgs boson coupling modifiers: \texorpdfstring{$\kappat$}{kt} vs. \texorpdfstring{$\cg$}{cg} and \texorpdfstring{$\kappat$}{kt} vs. \texorpdfstring{$\kappab$}{kb}}

\begin{figure}[!hbt]
  \begin{center}
    \includegraphics[width=\cmsFigWidth]{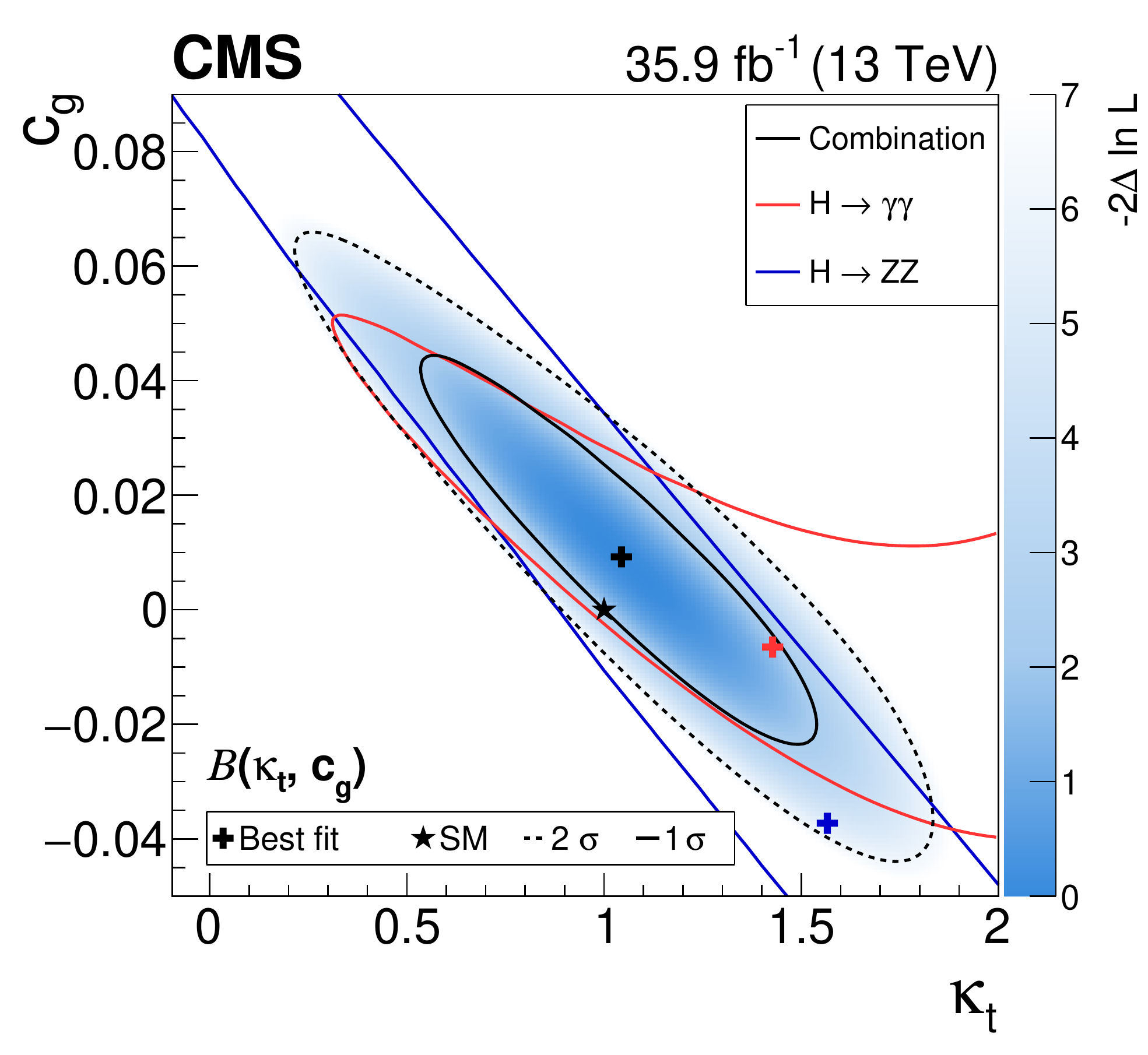}
    \includegraphics[width=\cmsFigWidth]{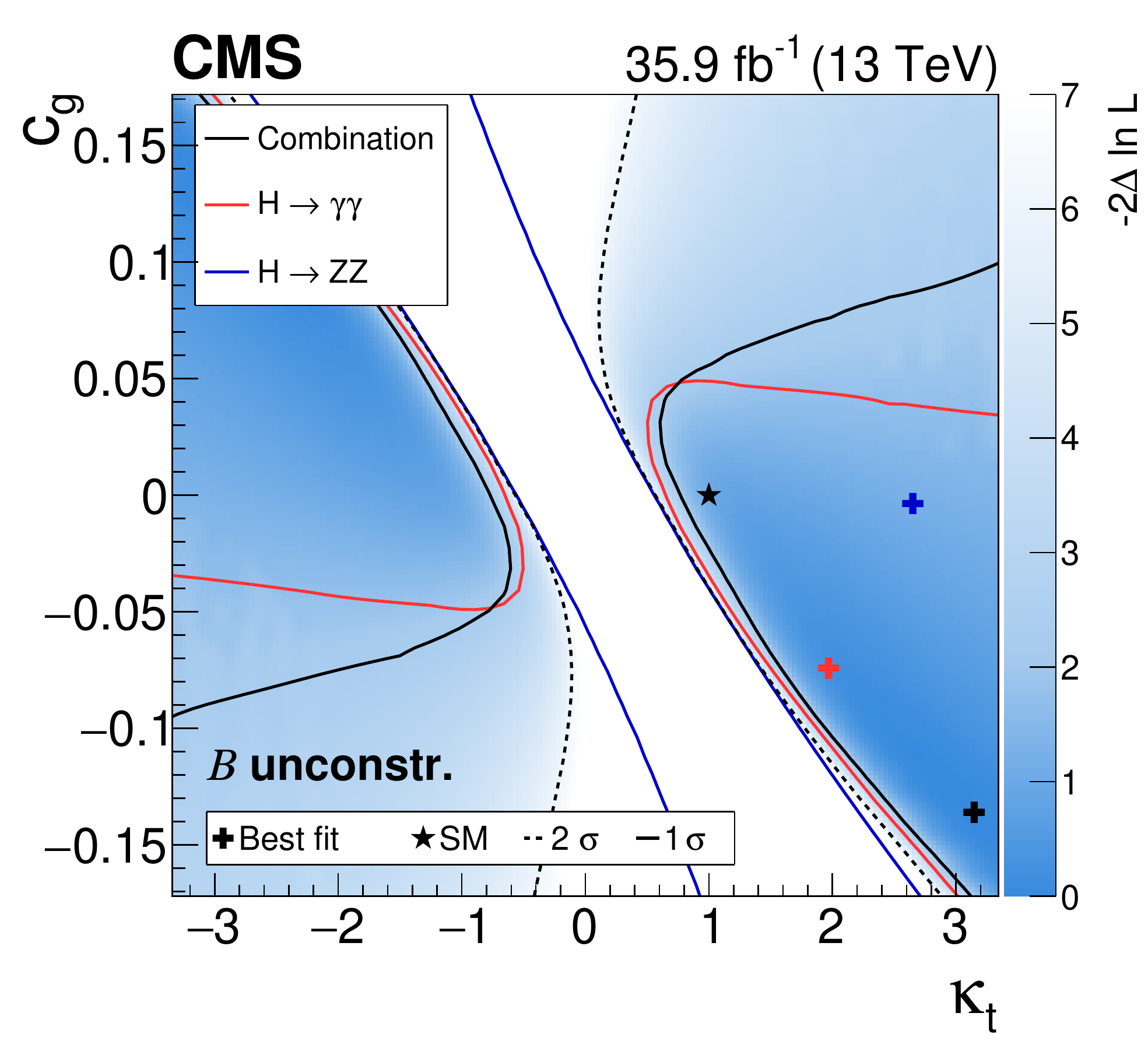}
        \caption{
        Simultaneous fit to data for $\kappat$ and $\cg$, assuming a coupling dependence of the branching fractions (\cmsLeft) and the branching fractions implemented as nuisance parameters with no prior constraint (\cmsRight).
                The one standard deviation contour is drawn for the combination ($\hgg$, $\hzz$, and $\hbb$), the $\hgg$ channel, and the $\hzz$ channel in black, red, and blue, respectively.
                For the combination the two standard deviation contour is drawn as a black dashed line, and the shading indicates the negative log-likelihood, with the scale shown on the right hand side of the plots.
        }
    \label{fig:scans_kappatkappag_nominal}
  \end{center}
\end{figure}

The fits are repeated in a way analogous to that of Section~\ref{sec:ResultsKappabKappac} but with $\kappat$, $\cg$, and $\kappab$, the coefficients of the dimension-6 operators added to the SM Lagrangian, as the parameters of the fit, using the parametrization obtained from Refs.~\cite{Grazzini:2017szg,Grazzini:2016paz}.
The combined log-likelihood scan for $\kappat$ vs. $\cg$, assuming branching fractions that depend on the couplings, is shown in Fig.~\ref{fig:scans_kappatkappag_nominal} (\cmsLeft).
The normalization of the spectrum is, by construction, equal to the SM normalization for the set of coefficients satisfying $12 \cg + \kappat \simeq 1$.
The shape of the parametrized $\pth$ spectrum $s$ is calculated by normalizing the differential cross section to $1$:
\begin{linenomath*}
\begin{equation}
    s_i(\kappat, \cg) =
        \frac
            {\sigma_i(\kappat, \cg)}
            {\sum_j \sigma_j(\kappat, \cg)},
\end{equation}
\end{linenomath*}
where $\sigma_i$ is the parametrization in bin $i$.
Inserting the expected parabolic dependence of $\sigma_i(\kappat, \cg)$ reveals that the shape of the parametrization for $\kappat$/$\cg$ variations becomes a function of the ratio of the two couplings, $s_i({\cg}/{\kappat})$.
Thus the dependence of the likelihood on the radial distance $\sqrt{\smash[b]{\kappat^2+\cg^2}}$ stems from constraints on the overall normalization, whereas the dependence on the slope ${\cg}/{\kappat}$ stems from constraints on the shape of the distribution.
The dependence of the likelihood on the slope becomes apparent in Fig.~\ref{fig:scans_kappatkappag_nominal} (\cmsRight), where the branching fractions are implemented as nuisance parameters with no prior constraint in the fit.
Except at small values of the couplings, the constraint on the couplings comes from their ratio.
The two symmetric sets of contours are due to a symmetry of the parametrization under $(\kappat,\,\cg) \, \to \, (-\kappat,\,-\cg)$.
The constraint from the $\hgg$ channel individually is here slightly stronger than the combination; this effect, not observed in expected fits, stems from opposite deviations in the $\hgg$ and $\hzz$ $\pth$ spectra that cancel out in the combination.

\begin{figure}[hbt!]
  \begin{center}
    \includegraphics[width=\cmsFigWidth]{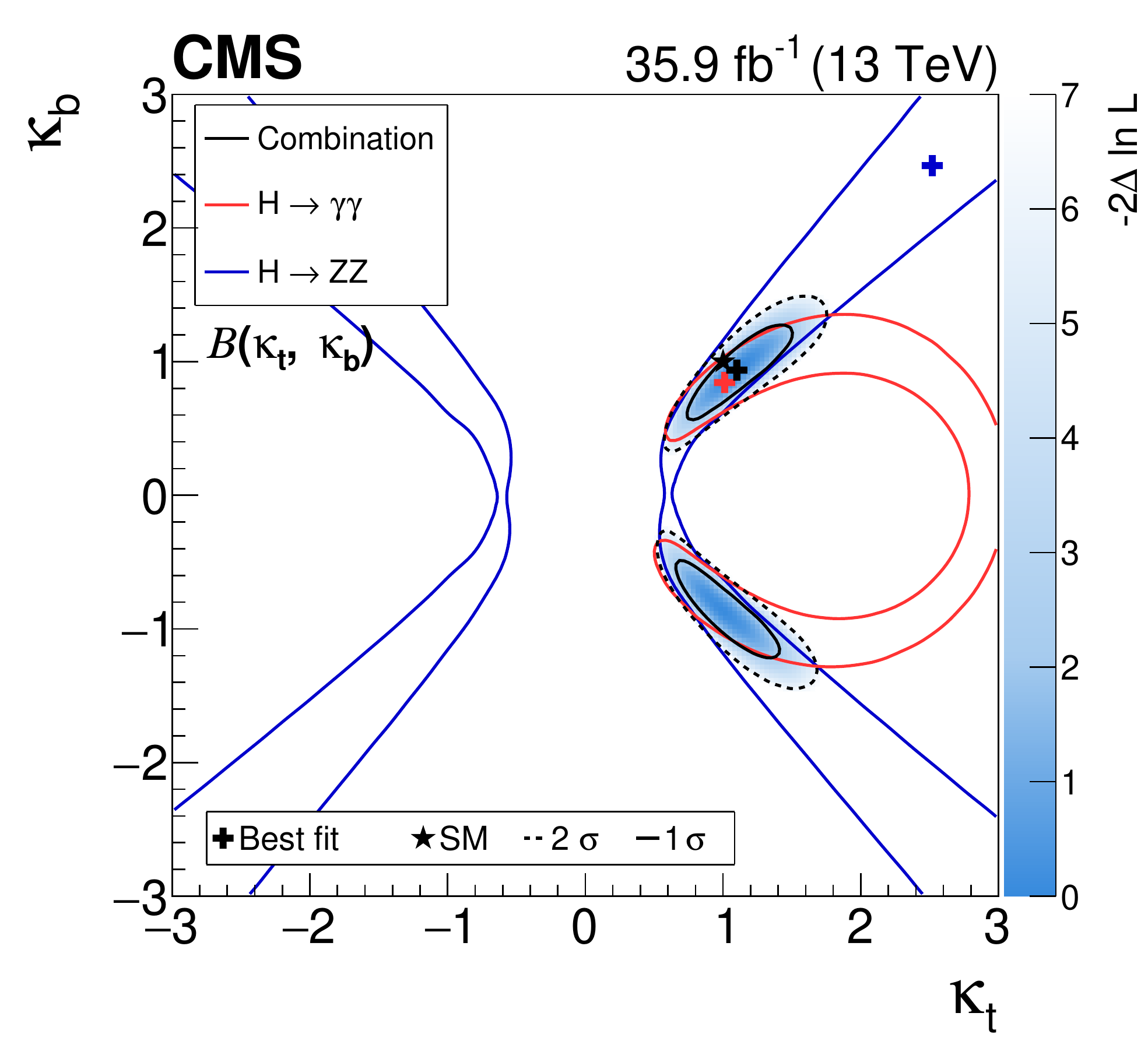}
    \includegraphics[width=\cmsFigWidth]{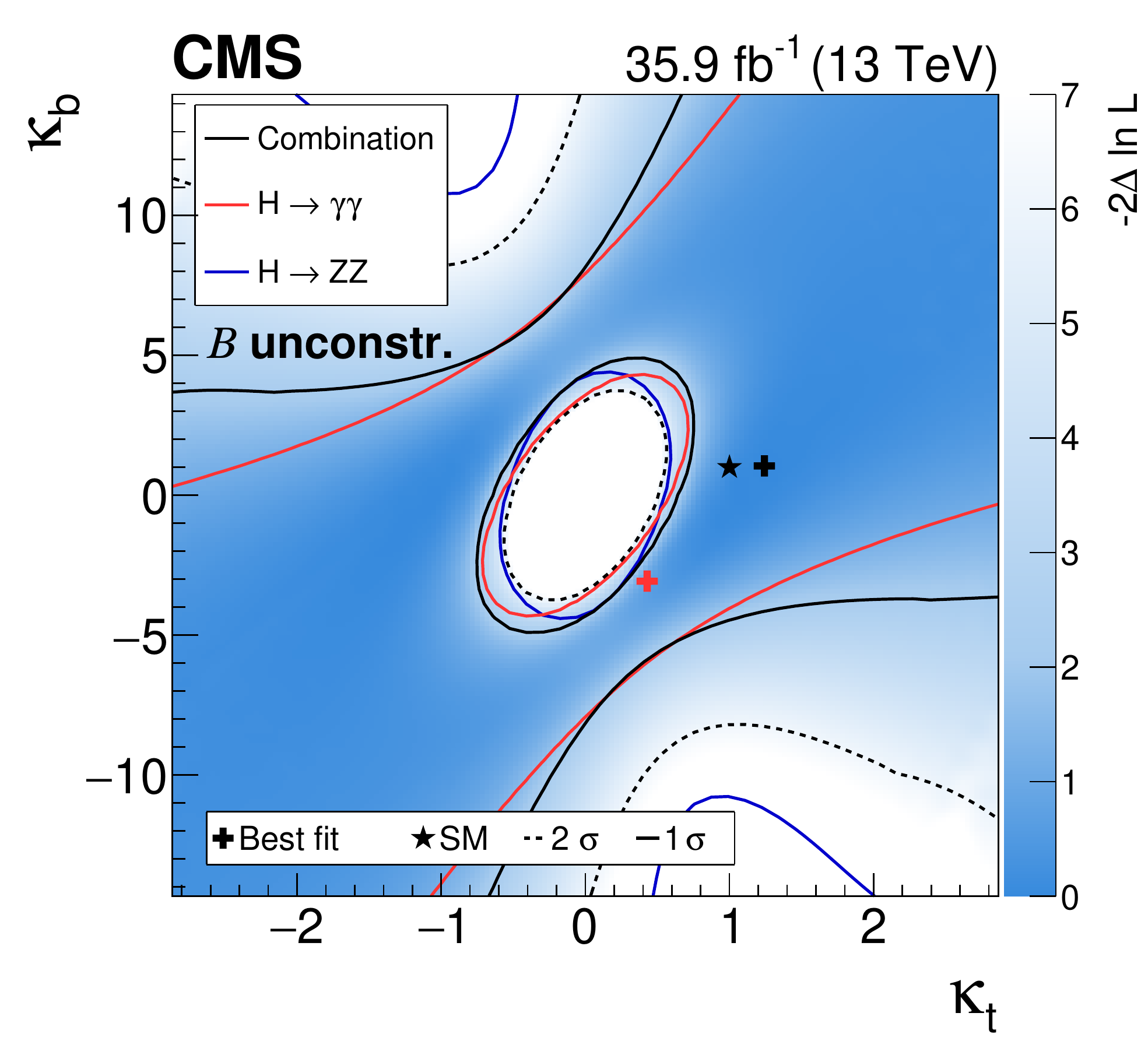}
        \caption{
        Simultaneous fit to data for $\kappat$ and $\kappab$, assuming a coupling dependence of the branching fractions (\cmsLeft) and the branching fractions implemented as nuisance parameters with no prior constraint (\cmsRight).
                The one standard deviation contour is drawn for the combination ($\hgg$, $\hzz$, and $\hbb$), the $\hgg$ channel, and the $\hzz$ channel in black, red, and blue, respectively.
                For the combination the two standard deviation contour is drawn as a black dashed line, and the shading indicates the negative log-likelihood, with the scale shown on the right hand side of the plots.
        }
    \label{fig:scans_kappatkappab_rawInput}
  \end{center}
\end{figure}

Figure~\ref{fig:scans_kappatkappab_rawInput} (\cmsLeft) shows the combined log-likelihood scan as a function of $\kappat$ and $\kappab$, with branching fractions scaling appropriately with the coupling modifiers and Fig.~\ref{fig:scans_kappatkappab_rawInput} (\cmsRight) with the branching fractions implemented as nuisance parameters with no prior constraint.
As the $\hgg$ branching fraction depends linearly on $\kappat$, the constraints on the $\hgg$ channel and the combination in Fig.~\ref{fig:scans_kappatkappab_rawInput} (\cmsLeft) are not symmetric with respect to the $\kappat$ axis.
For the branching fractions implemented as nuisance parameters with no prior constraint, the parametrization is symmetric under $(\kappat,\,\kappab) \, \to \, (-\kappat,\,-\kappab)$, which explains the observed symmetry in Fig.~\ref{fig:scans_kappatkappab_rawInput} (\cmsRight).

\section{Summary}

A combination of differential cross sections for the Higgs boson transverse momentum $\pth$, the number of jets, the rapidity of the Higgs boson, and the $\pt$ of the leading jet has been presented, using proton-proton collision data collected at $\sqrt{s}=13$\TeV with the CMS detector, corresponding to an integrated luminosity of $35.9$\fbinv.
The spectra obtained are based on data from the $\hgg$, $\hzz$, and $\hbb$ decay channels.
The precision of the combined measurement of the differential cross section of $\pth$ is improved by about 15\% with respect to the $\hgg$ channel alone.
The improvement is larger in the low-$\pth$ region than in the high-$\pth$ tails.
No significant deviations from the standard model are observed in any differential distribution.
Additionally, the total cross section for Higgs boson production based on a combination of the $\hgg$ and $\hzz$ channels is measured to be $61.1   \pm 6.0 \stat   \pm 3.7 \syst  $\pb.

The spectra obtained are interpreted in the \kappaframework~\cite{LHCHXSWG:YR3}, in which simultaneous variations of $\kappab$ and $\kappac$, $\kappat$ and $\kappab$, and $\kappat$ and the anomalous direct coupling to the gluon field $\cg$ are fitted to the $\pth$ spectra.
The limits obtained for the individual couplings are $\kappabLeftObserved < \kappab < \kappabRightObserved$ and $\kappacLeftObserved < \kappac < \kappacRightObserved$ at 95\% confidence level, assuming the branching fractions scale with the Higgs boson couplings following the standard model prediction.
For the charm coupling $\kappac$ in particular, these bounds are comparable with those obtained from direct searches with charm quarks in the final state.

\begin{acknowledgments}
We thank Fady Bishara, Ulrich Haisch, Pier Francesco Monni, Emanuele Re, Massimiliano Grazzini, Agnieszka Ilnicka, Michael Spira, and Marius Wiesemann for guidance regarding their predictions of the Higgs boson transverse momentum spectra.
We congratulate our colleagues in the CERN accelerator departments for the excellent performance of the LHC and thank the technical and administrative staffs at CERN and at other CMS institutes for their contributions to the success of the CMS effort. In addition, we gratefully acknowledge the computing centres and personnel of the Worldwide LHC Computing Grid for delivering so effectively the computing infrastructure essential to our analyses. Finally, we acknowledge the enduring support for the construction and operation of the LHC and the CMS detector provided by the following funding agencies: BMBWF and FWF (Austria); FNRS and FWO (Belgium); CNPq, CAPES, FAPERJ, FAPERGS, and FAPESP (Brazil); MES (Bulgaria); CERN; CAS, MoST, and NSFC (China); COLCIENCIAS (Colombia); MSES and CSF (Croatia); RPF (Cyprus); SENESCYT (Ecuador); MoER, ERC IUT, and ERDF (Estonia); Academy of Finland, MEC, and HIP (Finland); CEA and CNRS/IN2P3 (France); BMBF, DFG, and HGF (Germany); GSRT (Greece); NKFIA (Hungary); DAE and DST (India); IPM (Iran); SFI (Ireland); INFN (Italy); MSIP and NRF (Republic of Korea); MES (Latvia); LAS (Lithuania); MOE and UM (Malaysia); BUAP, CINVESTAV, CONACYT, LNS, SEP, and UASLP-FAI (Mexico); MOS (Montenegro); MBIE (New Zealand); PAEC (Pakistan); MSHE and NSC (Poland); FCT (Portugal); JINR (Dubna); MON, RosAtom, RAS, RFBR, and NRC KI (Russia); MESTD (Serbia); SEIDI, CPAN, PCTI, and FEDER (Spain); MOSTR (Sri Lanka); Swiss Funding Agencies (Switzerland); MST (Taipei); ThEPCenter, IPST, STAR, and NSTDA (Thailand); TUBITAK and TAEK (Turkey); NASU and SFFR (Ukraine); STFC (United Kingdom); DOE and NSF (USA).

\hyphenation{Rachada-pisek} Individuals have received support from the Marie-Curie programme and the European Research Council and Horizon 2020 Grant, contract No. 675440 (European Union); the Leventis Foundation; the A.P.\ Sloan Foundation; the Alexander von Humboldt Foundation; the Belgian Federal Science Policy Office; the Fonds pour la Formation \`a la Recherche dans l'Industrie et dans l'Agriculture (FRIA-Belgium); the Agentschap voor Innovatie door Wetenschap en Technologie (IWT-Belgium); the F.R.S.-FNRS and FWO (Belgium) under the ``Excellence of Science -- EOS" -- be.h project n.\ 30820817; the Ministry of Education, Youth and Sports (MEYS) of the Czech Republic; the Lend\"ulet (``Momentum") Programme and the J\'anos Bolyai Research Scholarship of the Hungarian Academy of Sciences, the New National Excellence Program \'UNKP, the NKFIA research grants 123842, 123959, 124845, 124850, and 125105 (Hungary); the Council of Science and Industrial Research, India; the HOMING PLUS programme of the Foundation for Polish Science, cofinanced from European Union, Regional Development Fund, the Mobility Plus programme of the Ministry of Science and Higher Education, the National Science Center (Poland), contracts Harmonia 2014/14/M/ST2/00428, Opus 2014/13/B/ST2/02543, 2014/15/B/ST2/03998, and 2015/19/B/ST2/02861, Sonata-bis 2012/07/E/ST2/01406; the National Priorities Research Program by Qatar National Research Fund; the Programa Estatal de Fomento de la Investigaci{\'o}n Cient{\'i}fica y T{\'e}cnica de Excelencia Mar\'{\i}a de Maeztu, grant MDM-2015-0509 and the Programa Severo Ochoa del Principado de Asturias; the Thalis and Aristeia programmes cofinanced by EU-ESF and the Greek NSRF; the Rachadapisek Sompot Fund for Postdoctoral Fellowship, Chulalongkorn University and the Chulalongkorn Academic into Its 2nd Century Project Advancement Project (Thailand); the Welch Foundation, contract C-1845; and the Weston Havens Foundation (USA).
\end{acknowledgments}

\bibliography{auto_generated}
\clearpage
\numberwithin{equation}{section}
\numberwithin{table}{section}
\numberwithin{figure}{section}
\appendix

\section{Tables for the differential cross section measurements}
\label{sec:tables}

Tables~\ref{tab:numbers_pth_smH}--\ref{tab:numbers_ptjet} show the measured differential cross sections for the considered observables.

\begin{table*}[!ht]
    \centering
    \topcaption{
        Differential cross sections (pb/\GeVns{}) for the observable $\pth$.
        }
    \label{tab:numbers_pth_smH}
    \setlength{\tabcolsep}{3pt}
    \renewcommand*{\arraystretch}{1.4}
    \cmsTableForced{
    \begin{tabular}{lccccccccc}
$\pth$ (\GeVns{})                         & 0--15                      & 15--30                   & 30--45                     & 45--80                      & 80--120                      & 120--200                      & 200--350                                                & 350--600                                                              & $>$600                                                               \\[\cmsTabSkip]
\hline
$\hgg$
    & $1.0 \, {{}}^{+0.3}_{-0.3}$ & $1 \, {{}}^{+0.3}_{-0.3}$ & $0.5 \, {{}}^{+0.2}_{-0.2}$ & $0.3 \, {{}}^{+0.1}_{-0.1}$  & $0.1 \, {{}}^{+0.05}_{-0.05}$ & $0.03 \, {{}}^{+0.01}_{-0.01}$ & $0.01 \, {{}}^{+2.8 \times 10^{-3}}_{-2.5 \times 10^{-3}}$ & $-3.4 \times 10^{-5} \, {{}}^{+3.8 \times 10^{-4}}_{-3.1 \times 10^{-4}}$ & $-1.9 \times 10^{-4} \, {{}}^{+2.4 \times 10^{-4}}_{-2.4 \times 10^{-4}}$ \\
$\hzz$
    & $0.7 \, {{}}^{+0.3}_{-0.3}$ & $1 \, {{}}^{+0.4}_{-0.3}$ & \multicolumn{2}{l}{$0.4 \, {{}}^{+0.1}_{-0.1}$}         & \multicolumn{2}{l}{$0.08 \, {{}}^{+0.03}_{-0.02}$}          & \multicolumn{3}{l}{$3.3 \times 10^{-4} \, {{}}^{+2.6 \times 10^{-3}}_{-2.6 \times 10^{-3}}$} \\
$\hbb$
    & \multicolumn{7}{l}{\textit{None}} & $9.6 \times 10^{-4} \, {{}}^{+1.2 \times 10^{-3}}_{-1.2 \times 10^{-3}}$  & $1.1 \times 10^{-4} \, {{}}^{+1.2 \times 10^{-4}}_{-1.1 \times 10^{-4}}$  \\
Comb.
    & $0.8 \, {{}}^{+0.2}_{-0.2}$ & $1 \, {{}}^{+0.2}_{-0.3}$ & $0.6 \, {{}}^{+0.2}_{-0.2}$ & $0.3 \, {{}}^{+0.1}_{-0.09}$ & $0.1 \, {{}}^{+0.05}_{-0.04}$ & $0.03 \, {{}}^{+0.01}_{-0.01}$ & $0.01 \, {{}}^{+2.6 \times 10^{-3}}_{-2.4 \times 10^{-3}}$ & $-2.8 \times 10^{-6} \, {{}}^{+3.7 \times 10^{-4}}_{-2.8 \times 10^{-4}}$ & $5.8 \times 10^{-5} \, {{}}^{+1.0 \times 10^{-4}}_{-1.0 \times 10^{-4}}$
    \end{tabular}
    }
    \end{table*}

\begin{table*}[!ht]
    \centering
    \topcaption{
        Differential cross sections of gluon fusion ($\ggh$) (pb/GeV) for the observable $\pth$, with non-\ggh production modes fixed to their SM prediction.
        }
    \label{tab:numbers_pth_ggH}
    \setlength{\tabcolsep}{3pt}
    \renewcommand*{\arraystretch}{1.4}
    \cmsTableForced{
    \begin{tabular}{lccccccccc}
$\pth$ (\GeVns{}) & 0--15                      & 15--30                   & 30--45                     & 45--80                      & 80--120                      & 120--200                      & 200--350                                                             & 350--600                                                              & $>$600                                                              \\[\cmsTabSkip]
\hline
Comb.  & $0.8 \, {{}}^{+0.2}_{-0.2}$ & $1 \, {{}}^{+0.2}_{-0.3}$ & $0.5 \, {{}}^{+0.2}_{-0.2}$ & $0.2 \, {{}}^{+0.1}_{-0.09}$ & $0.1 \, {{}}^{+0.05}_{-0.04}$ & $0.02 \, {{}}^{+0.01}_{-0.01}$ & $8.3 \times 10^{-3} \, {{}}^{+2.6 \times 10^{-3}}_{-2.4 \times 10^{-3}}$ & $-1.6 \times 10^{-4} \, {{}}^{+3.4 \times 10^{-4}}_{-2.6 \times 10^{-4}}$ & $3.5 \times 10^{-5} \, {{}}^{+5.8 \times 10^{-5}}_{-5.7 \times 10^{-5}}$
    \end{tabular}
    }
    \end{table*}

\begin{table*}[!ht]
    \centering
    \topcaption{
        Differential cross sections (pb) for the observable $\njets$.
        }
    \label{tab:numbers_njets}
    \renewcommand*{\arraystretch}{1.4}
    \begin{tabular}{lccccc}
    $\njets$
        & 0                                                                & 1                                                                & 2                                                                 & 3                                                                & $\ge$4                                                            \\[\cmsTabSkip]
        \hline
    $\hgg$
        & $50 \, {{}}^{+8.5}_{-8.1}$ & $14 \, {{}}^{+5.1}_{-4.9}$ & $4.8 \times 10^{-1} \, {{}}^{+2.7}_{-2.7}$ & $3.1 \, {{}}^{+2.0}_{-2.0}$ & $1.3 \, {{}}^{+8.8 \times 10^{-1}}_{-9.3 \times 10^{-1}}$ \\
    $\hzz$
        & $41 \, {{}}^{+9.1}_{-8.0}$ & $8.7 \, {{}}^{+5.2}_{-4.3}$ & $6.9 \, {{}}^{+3.7}_{-3.0}$  & \multicolumn{2}{l}{$1.2 \, {{}}^{+2.1}_{-2.1}$}                                                 \\
    Combination
        & $47 \, {{}}^{+6.2}_{-6.4}$ & $11 \, {{}}^{+3.7}_{-3.4}$ & $3.5 \, {{}}^{+1.9}_{-1.7}$  & $1.8 \, {{}}^{+1.7}_{-1.5}$ & $1.2 \, {{}}^{+8.3 \times 10^{-1}}_{-8.8 \times 10^{-1}}$
    \end{tabular}
    \end{table*}

\begin{table*}[!ht]
    \centering
    \topcaption{
        Differential cross sections (pb) for the observable $\absy$.
        }
    \label{tab:numbers_absy}
    \renewcommand*{\arraystretch}{1.4}
    \begin{tabular}{lcccccc}
    $\absy$
        & 0--0.15                                                           & 0.15--0.3                                                         & 0.3--0.6                                                          & 0.6--0.9                                                          & 0.9--1.2                                                          & 1.2--2.5                                                          \\[\cmsTabSkip]
        \hline
    $\hgg$
        & $42 \, {{}}^{+11}_{-11}$ & $39 \, {{}}^{+12}_{-11}$ & $31 \, {{}}^{+9.0}_{-7.5}$ & $28 \, {{}}^{+9.1}_{-8.7}$ & $24 \, {{}}^{+12}_{-10}$ & $18 \, {{}}^{+7.4}_{-7.2}$ \\
    $\hzz$
        & $39 \, {{}}^{+17}_{-14}$ & $35 \, {{}}^{+18}_{-14}$ & $34 \, {{}}^{+11}_{-9.8}$ & $45 \, {{}}^{+13}_{-11}$ & $13 \, {{}}^{+8.9}_{-6.8}$ & $13 \, {{}}^{+6.7}_{-5.4}$ \\
    Combination
        & $41 \, {{}}^{+9.1}_{-8.9}$ & $38 \, {{}}^{+9.7}_{-9.2}$ & $32 \, {{}}^{+7.0}_{-6.0}$ & $35 \, {{}}^{+7.1}_{-6.6}$ & $17 \, {{}}^{+7.4}_{-6.5}$ & $15 \, {{}}^{+5.1}_{-4.7}$
    \end{tabular}
    \end{table*}

\begin{table*}[!ht]
    \centering
    \topcaption{
        Differential cross sections (pb/GeV) for the observable $\ptjet$.
        }
    \label{tab:numbers_ptjet}
    \renewcommand*{\arraystretch}{1.4}
    \cmsTableForced{
    \begin{tabular}{lccccc}
    $\ptjet$ (\GeV)
        & 30--55                                                               & 55--95                                                               & 95--120                                                              & 120--200                                                              & $>$200                                                              \\[\cmsTabSkip]
        \hline
    $\hgg$
        & $1.6 \times 10^{-1} \, {{}}^{+2.0 \times 10^{-1}}_{-2.1 \times 10^{-1}}$ & $2.0 \times 10^{-1} \, {{}}^{+9.2 \times 10^{-2}}_{-9.3 \times 10^{-2}}$ & $1.3 \times 10^{-1} \, {{}}^{+9.5 \times 10^{-2}}_{-9.2 \times 10^{-2}}$ & $1.5 \times 10^{-5} \, {{}}^{+1.8 \times 10^{-2}}_{-1.7 \times 10^{-2}}$  & $2.9 \times 10^{-2} \, {{}}^{+9.1 \times 10^{-3}}_{-9.2 \times 10^{-3}}$ \\
    $\hzz$
        & $4.8 \times 10^{-1} \, {{}}^{+2.4 \times 10^{-1}}_{-2.0 \times 10^{-1}}$ & $7.7 \times 10^{-2} \, {{}}^{+8.8 \times 10^{-2}}_{-6.9 \times 10^{-2}}$ & \multicolumn{3}{l}{$8.0 \times 10^{-2} \, {{}}^{+5.9 \times 10^{-2}}_{-4.4 \times 10^{-2}}$} \\
    Combination
        & $3.2 \times 10^{-1} \, {{}}^{+1.4 \times 10^{-1}}_{-1.3 \times 10^{-1}}$ & $1.3 \times 10^{-1} \, {{}}^{+7.7 \times 10^{-2}}_{-6.1 \times 10^{-2}}$ & $1.1 \times 10^{-1} \, {{}}^{+8.4 \times 10^{-2}}_{-8.1 \times 10^{-2}}$ & $-4.2 \times 10^{-3} \, {{}}^{+1.7 \times 10^{-2}}_{-1.6 \times 10^{-2}}$ & $2.7 \times 10^{-2} \, {{}}^{+8.7 \times 10^{-3}}_{-8.9 \times 10^{-3}}$
    \end{tabular}
    }
    \end{table*}

\clearpage
\section{Correlation matrices for the combinations of differential observables}
\label{sec:binToBinCorrelationMatrices}

\begin{figure}[hbtp]
  \begin{center}
    \includegraphics[width=\cmsFigWidth]{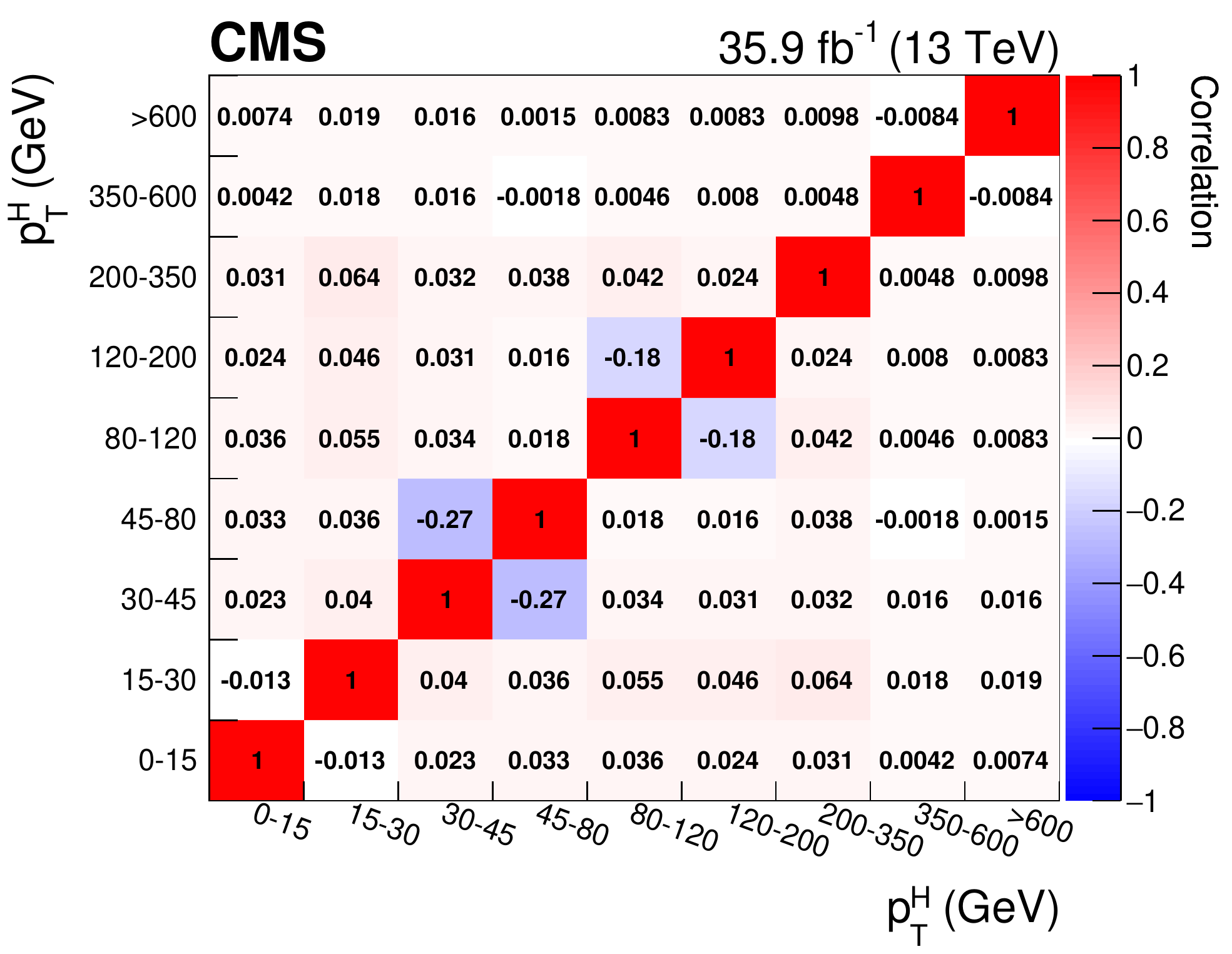}
    \includegraphics[width=\cmsFigWidth]{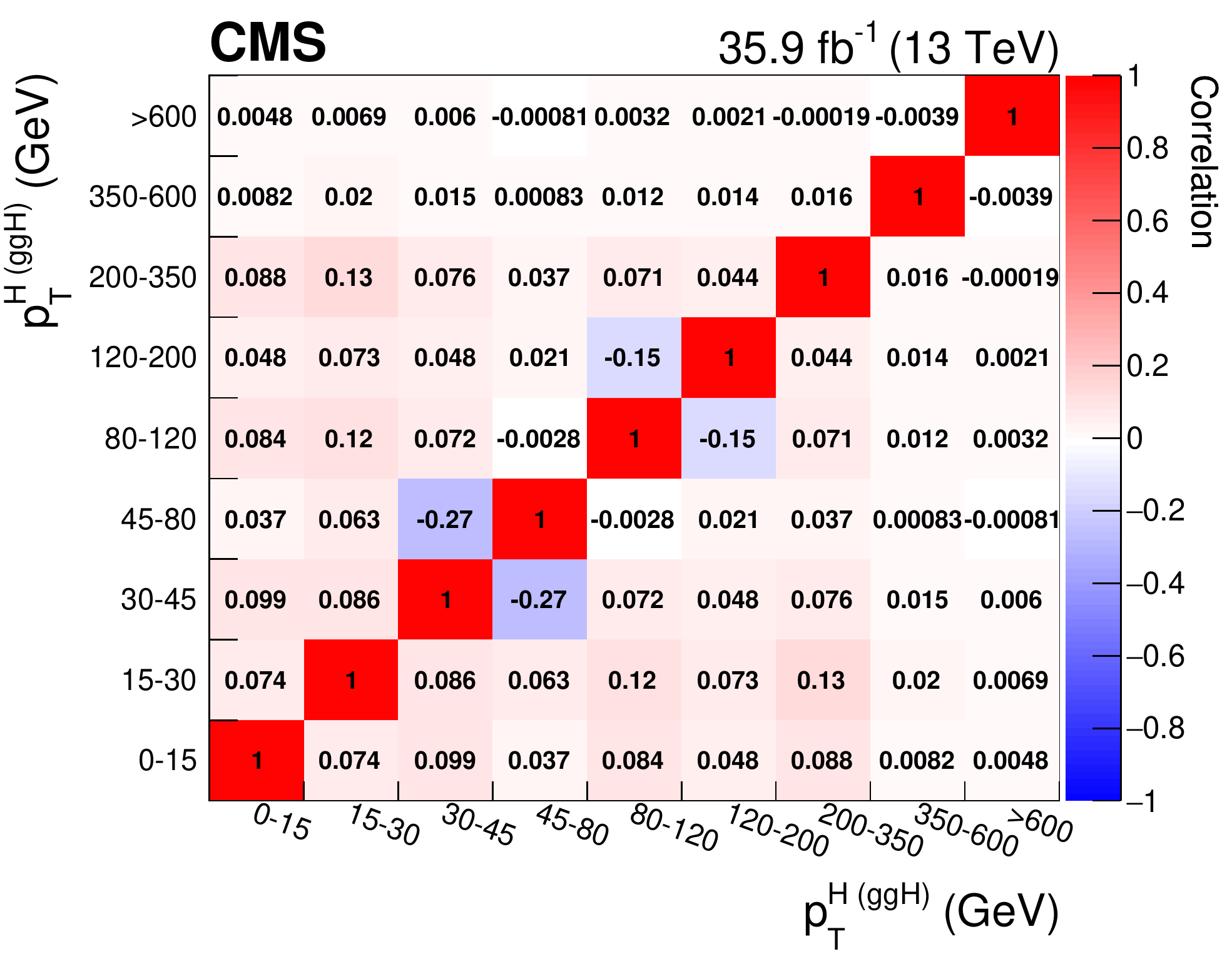}
    \caption{
         Bin-to-bin correlation matrix of the $\pth$ spectrum (\cmsLeft) and of the $\pth$ spectrum of gluon fusion ($\ggh$), where the non-\ggh contributions are fixed to the SM expectation (\cmsRight).
        }
    \label{fig:corrMat_pth}
  \end{center}
\end{figure}

\begin{figure}[hbtp]
  \begin{center}
    \includegraphics[width=\cmsFigWidth]{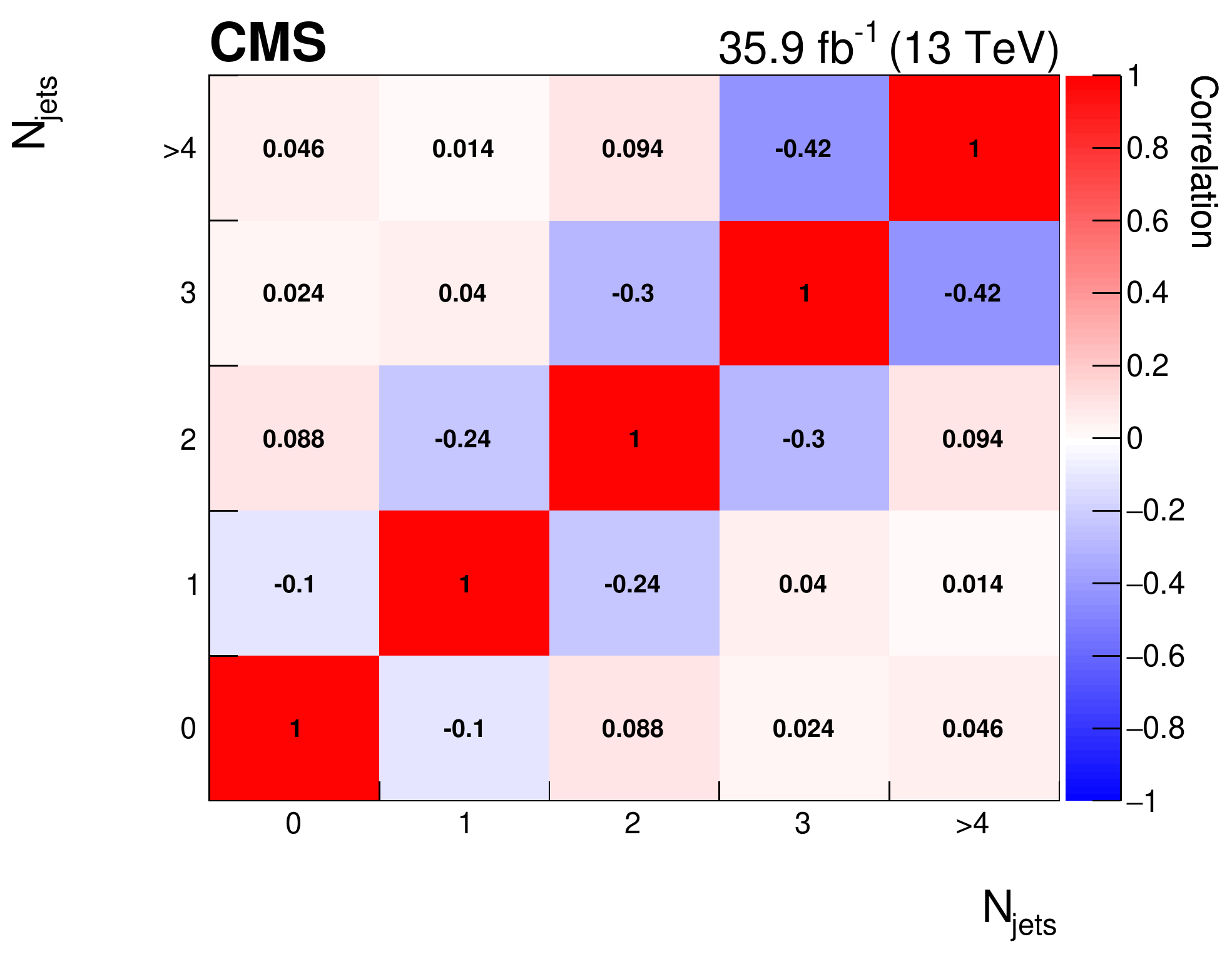}
    \caption{
        Bin-to-bin correlation matrix of the $\njets$ spectrum.
        }
    \label{fig:corrMat_njets}
  \end{center}
\end{figure}

\begin{figure}[hbtp]
  \begin{center}
    \includegraphics[width=\cmsFigWidth]{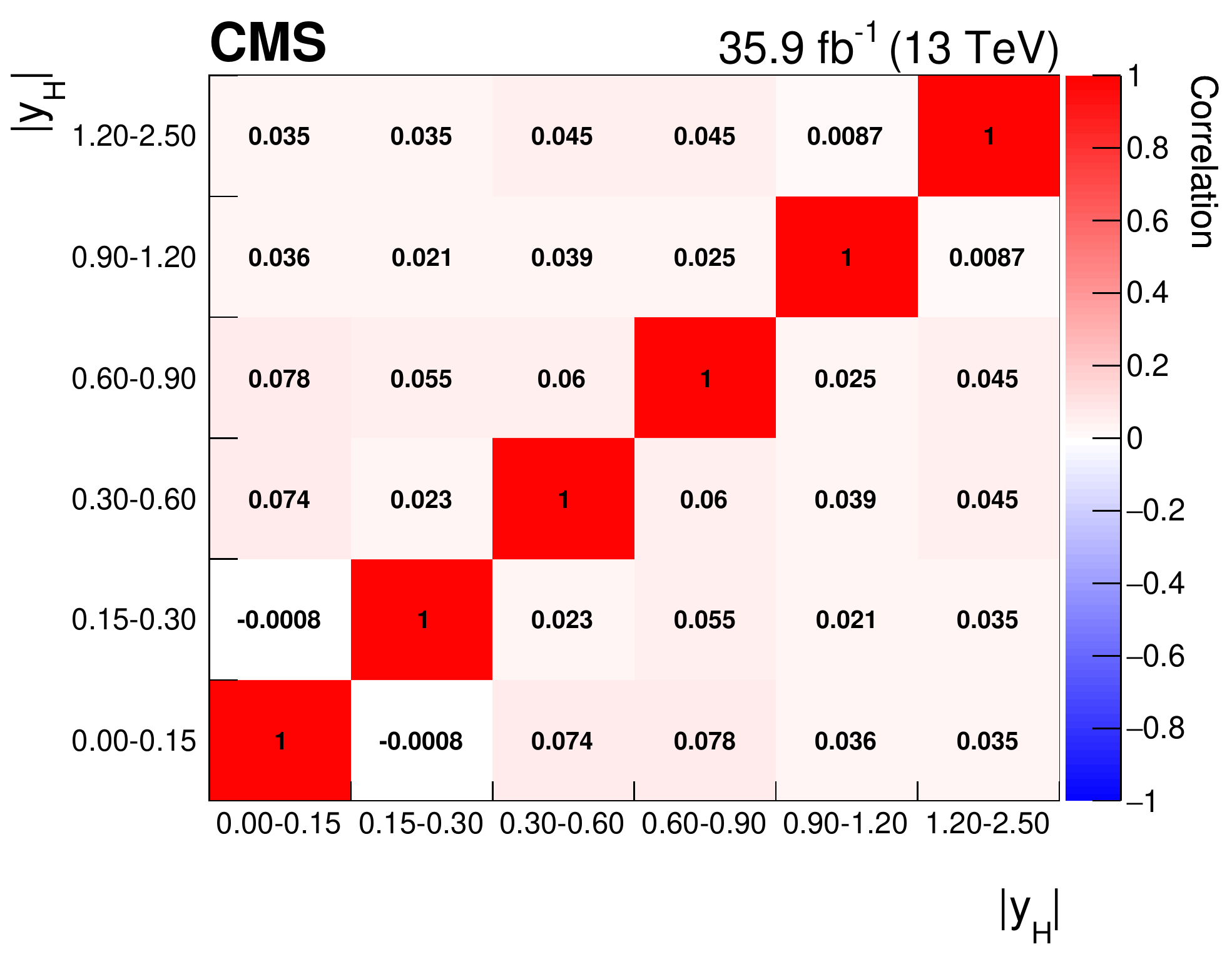}
    \caption{
        Bin-to-bin correlation matrix of the $\absy$ spectrum.
        }
    \label{fig:corrMat_absy}
  \end{center}
\end{figure}

\begin{figure}[hbtp]
  \begin{center}
    \includegraphics[width=\cmsFigWidth]{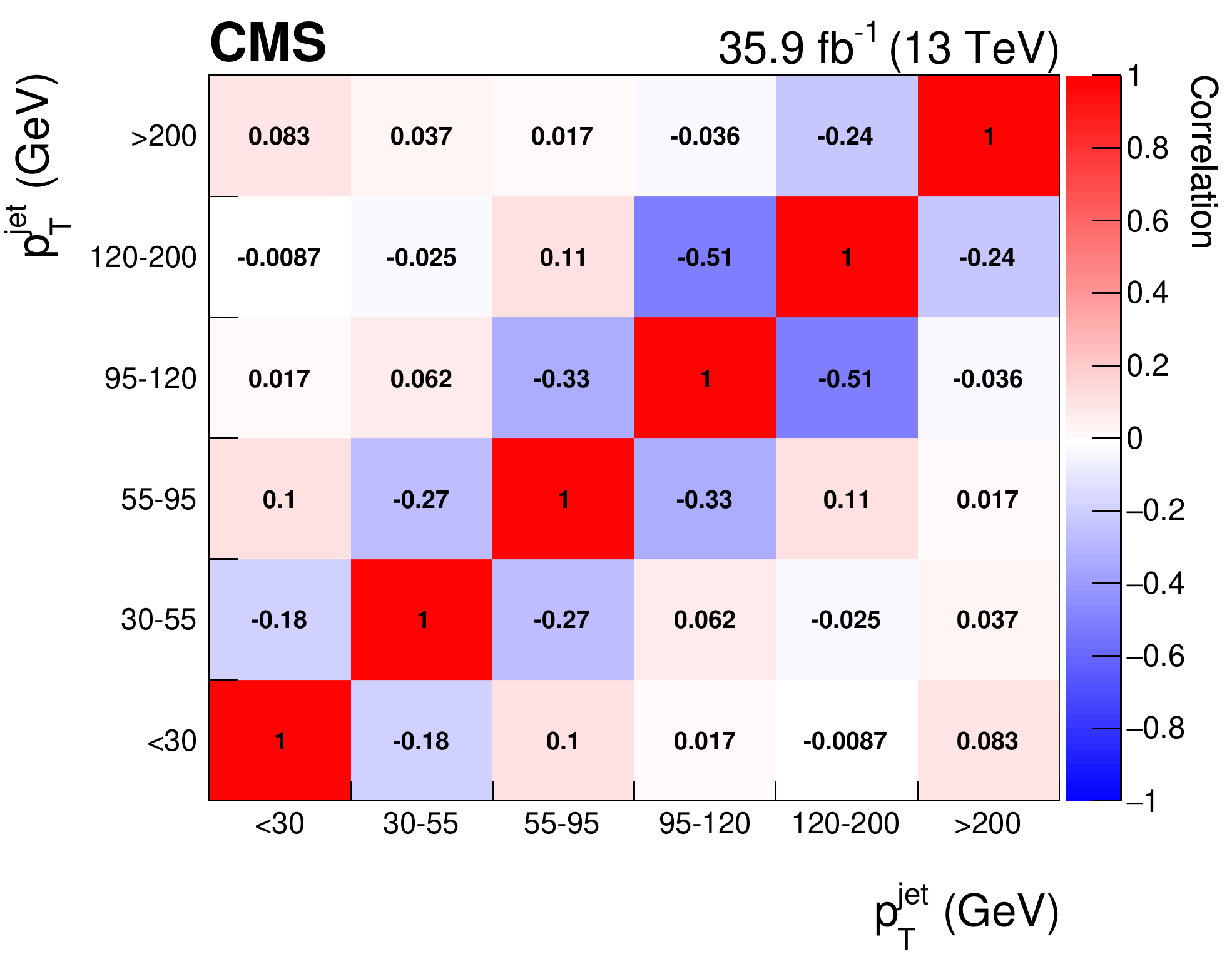}
    \caption{
        Bin-to-bin correlation matrix of the $\ptjet$ spectrum.
        }
    \label{fig:corrMat_ptjet}
  \end{center}
\end{figure}

Figs.~\ref{fig:corrMat_pth}--\ref{fig:corrMat_ptjet} show the correlation matrices for the considered observables.

 \cleardoublepage \section{The CMS Collaboration \label{app:collab}}\begin{sloppypar}\hyphenpenalty=5000\widowpenalty=500\clubpenalty=5000\vskip\cmsinstskip
\textbf{Yerevan Physics Institute, Yerevan, Armenia}\\*[0pt]
A.M.~Sirunyan, A.~Tumasyan
\vskip\cmsinstskip
\textbf{Institut f\"{u}r Hochenergiephysik, Wien, Austria}\\*[0pt]
W.~Adam, F.~Ambrogi, E.~Asilar, T.~Bergauer, J.~Brandstetter, M.~Dragicevic, J.~Er\"{o}, A.~Escalante~Del~Valle, M.~Flechl, R.~Fr\"{u}hwirth\cmsAuthorMark{1}, V.M.~Ghete, J.~Hrubec, M.~Jeitler\cmsAuthorMark{1}, N.~Krammer, I.~Kr\"{a}tschmer, D.~Liko, T.~Madlener, I.~Mikulec, N.~Rad, H.~Rohringer, J.~Schieck\cmsAuthorMark{1}, R.~Sch\"{o}fbeck, M.~Spanring, D.~Spitzbart, A.~Taurok, W.~Waltenberger, J.~Wittmann, C.-E.~Wulz\cmsAuthorMark{1}, M.~Zarucki
% ...
\vskip\cmsinstskip
% ...
1:  Also at Vienna University of Technology, Vienna, Austria\\
% ...
\end{sloppypar}
\end{document}